\documentclass[twocolumn]{aastex62}
\usepackage{multirow}
\usepackage{times}
\usepackage{natbib}
\usepackage{graphicx,cancel}
\usepackage{booktabs}
\usepackage{tablefootnote}
\usepackage{amsmath}
\usepackage{float}
\usepackage{xspace}
\usepackage{verbatim} 
\usepackage{txfonts}

\bibpunct{(}{)}{;}{a}{}{,}

\def\pdot {\dot P}
\def\edot {\dot E}

\def\x{\times}
\def\E{\mathit{E}}

\def\nh{$N_{\rm H}$\xspace}

\def\msun{~M_{\odot}}

\def\ltsima{$\; \buildrel < \over \sim \;$}
\def\lsim{\lower.5ex\hbox{\ltsima}}
\def\gtsima{$\; \buildrel > \over \sim \;$}
\def\gsim{\lower.5ex\hbox{\gtsima}}
\newcommand{\band}[2]{$#1-#2$ keV\xspace}
\newcommand{\abs}[1]{\lvert #1 \rvert}
\newcommand{\coord}[8]{R.A.\,=\,${#1}^{\rm h} {#2}^{\rm m} {#3}^{\rm s}.{#4}$, dec.\,=\,${#5}^\circ {#6}' {#7}''.{#8}$} 
\def\psrx{PSR\,B0943$+$10\xspace}
\def\xmm{{\em XMM--Newton}\xspace}

\received{2018 September 03}
\revised{2018 November 30}
\accepted{2018 December 20}

\submitjournal{ApJ}

\shorttitle{Thermal emission and magnetic beaming in \psrx}
\shortauthors{Rigoselli et al.}

\begin{document}

\title{Thermal emission and magnetic beaming in the radio and X-ray mode-switching \psrx.}

\correspondingauthor{Michela Rigoselli}
\email{m.rigoselli@campus.unimib.it, michela.rigoselli@inaf.it}

\author{Michela Rigoselli}
\affiliation{INAF, Istituto di Astrofisica Spaziale e Fisica Cosmica Milano, via E.\ Bassini 15, I-20133 Milano, Italy}
\affiliation{Dipartimento di Fisica G. Occhialini, Universit\`a degli Studi di Milano Bicocca, Piazza della Scienza 3, I-20126 Milano, Italy}

\author{Sandro Mereghetti}
\affiliation{INAF, Istituto di Astrofisica Spaziale e Fisica Cosmica Milano, via E.\ Bassini 15, I-20133 Milano, Italy}

\author{Roberto Turolla}
\affiliation{Dipartimento di Fisica e Astronomia, Universit\`a di Padova, via F. Marzolo 8, I-35131 Padova, Italy}
\affiliation{MSSL-UCL, Holmbury St. Mary, Dorking, Surrey RH5 6NT, UK}

\author{Roberto Taverna}
\affiliation{Dipartimento di Fisica e Astronomia, Universit\`a di Padova, via F. Marzolo 8, I-35131 Padova, Italy}
\affiliation{Dipartimento di Matematica e Fisica, Universit\`a di Roma Tre, via della Vasca Navale 84, I-00146 Roma, Italy}

\author{Valery Suleimanov}
\affiliation{Institut fur Astronomie und Astrophysik, Sand 1, 72076 Tubingen, Germany}
\affiliation{Space Research Institute of the Russian Academy of Sciences, Profsoyuznaya Str. 84/32, Moscow 117997, Russia}
\affiliation{Kazan (Volga region) Federal University, Kremlevskaja str., 18, Kazan 420008, Russia}
 
\author{Alexander Y. Potekhin}
\affiliation{Ioffe Institute, Politekhnicheskaya 26, 194021, Saint Petersburg, Russia}

\begin{abstract}

\psrx is a mode-switching radio pulsar characterized by two emission modes with different radio and X-ray properties. Previous studies, based on simple combinations of blackbody and power law models, 
showed that its X-ray flux can be decomposed in a pulsed thermal plus an unpulsed non-thermal components. However, if \psrx is a nearly aligned rotator seen pole-on, as suggested by the radio data, it is difficult to reproduce the high observed pulsed fraction unless magnetic beaming is included.
In this work we reanalyze all the available X-ray observations of \psrx with simultaneous radio coverage, 
modeling its thermal emission with polar caps covered by a magnetized hydrogen atmosphere or with a condensed iron surface.
The condensed surface model provides good fits to the spectra of both pulsar modes, but, similarly to the blackbody, it  can not reproduce the observed pulse profiles, unless an additional power law with an \textit{ad hoc} modulation is added. 
Instead, the pulse profiles and phase-resolved spectra are well described using the  hydrogen atmosphere model to describe the polar cap emission, plus an unpulsed power law. 
For the X-ray brighter state (Q-mode) we obtain a best fit with a temperature $kT \sim 0.09$ keV, an emitting radius $R \sim 260$ m, a magnetic field consistent with the value of the dipole field of $4\times10^{12}$ G inferred from the timing parameters, and a small angle between the magnetic and spin axis, $\xi=5^{\circ}$. The corresponding parameters for the X-ray fainter state (B-mode) are  $kT \sim 0.08$ keV and $R \sim 170$ m.  
\end{abstract}

\keywords{pulsar: general -- pulsars: individual (\psrx) -- stars: neutron -- X-rays: stars}

\section{Introduction}

It is generally believed that the thermal components observed in the X-ray spectra of many old and middle-aged rotation-powered pulsars originate from limited regions of the star surface, typically the polar caps, kept hot by some heating process.
In fact, as it has been known from the neutron star cooling theory since 1960s \citep{tsu66}, at an age $\gtrsim 1$ Myr the surface of an isolated neutron star becomes too cool to emit significantly in the X-ray range (see, e.g., \citealt{pot15,pot18} for recent reviews and references).
On the  observational side, the evidence for the  small size of the hot regions comes from the fact that the thermal components are pulsed and from the dimensions of the emitting area inferred from spectral fits with blackbody or neutron star atmosphere models \citep[ and references therein]{pot14b}.

There are several processes that can alter  the thermal evolution of an isolated neutron star.
They can be broadly divided into two categories, according to where they act, either in the star interior or at the surface. For example, the decay of the magnetic field and plastic deformations of the crust driven by magnetic stresses can release a significant amount of energy that heats the outer layers of the star \citep{arr04,pon07,coo10,vig13,kam14}. 

The latter  processes are effective only for very strong magnetic fields ($B \gtrsim 10^{14}$ G), therefore they are mainly relevant for magnetars.
A process that can lead to localized hot spots  results from the  photon-pair cascades that develop in the pulsar magnetosphere, where  the backward accelerated particles collide with the star surface near  the polar caps and heat it \citep{rud75,aro79,har01,har02a}. 

A few important issues have to be taken into account when studying the  thermal emission from neutron stars with X-ray observations. Spectral fits with a single blackbody model are  generally used  as a starting point to distinguish thermal from non-thermal (power law) emission and provide a first estimate of the temperature and emitting radius (the latter is of course affected by the uncertainty on the star distance).

It was soon realized, however, that more realistic models should be used instead (see, e.g., \citealt{rom87,shi92,pav94}). Indeed much effort was put into developing atmosphere models which account for the effects of different chemical compositions, magnetization, and surface gravity (e.g., \citealt{pot14b} and references therein).  
The observed X-ray spectra and pulse profiles depend also on other parameters that are often poorly known, such as star mass and radius, and on the system geometry, i.e. the angles between the rotation and magnetic axis/line-of-sight (LOS). A further complication is the presence of additional X-ray emission due to non-thermal processes. In many objects thermal and non-thermal components contribute in a comparable way over the whole observed energy range and it is difficult to disentangle them. This can lead to correlated parameters and large uncertainties in the spectral fits and affects the energy-dependence of the pulsed fractions.

In this work we focus on the X-ray emission of the pulsar \psrx. Being a relatively old  pulsar at a distance of about 1 kpc, its X-ray flux is rather weak.
However, \psrx is particularly interesting because it is one of the best studied mode-switching pulsars \citep{sul84}.
Mode-switching pulsars alternate between two (or more) modes  characterized by different properties in the radio band, such as the average pulse profile, intensity, polarization and drift rate of subpulses \citep{cor13}.
\citet{her13} discovered that also the X-ray properties of \psrx change when it switches between the so called (radio) Bright and Quiet modes  (hereafter B- and Q-mode), with the X-ray flux a factor $\sim\!2.5$ times higher in the Q-mode.  More recently, \citet{mer16} found that pulsed thermal X-ray emission, associated to a small emitting area, is present during both modes. In these works blackbody components were used in the spectral fits, while here we explore more complex models.
Indeed, a physically realistic description of the pulsed thermal emission of \psrx , that remarkably varies in anticorrelation with the radio flux, is required in trying to understand the still unknown reasons for the mode-switching behavior in radio pulsars.

In section \ref{sec:psr} we first recall the results of previous X-ray observations of \psrx and then summarize the information on its geometry as derived from radio data.
In section \ref{sec:model} we illustrate the approach we used to compute the phase-dependent spectrum emitted by the pulsar.
We then describe the  data analysis (section \ref{sec:obs})  and the results obtained when the thermal emission is modeled with a simple blackbody (section \ref{sec:bb}), a magnetized hydrogen atmosphere (section \ref{sec:atmo}) and a  magnetized condensed surface of iron (section \ref{sec:solid}).  The results are discussed in section \ref{sec:disc}.

\section{PSR B0943+10} 
\label{sec:psr}

The timing parameters of \psrx  (spin period $P=1.1$ s and $\pdot = 3.5\times 10^{-15}$ s~s$^{-1}$) give  a characteristic age $\tau = P/(2\pdot) = 5$ Myr,  a dipolar magnetic field at the poles $B_d = 4\times10^{12}$ G, and a rate of rotational energy losses $\edot =10^{32}$  erg s$^{-1}$.  Its dispersion measure is  $\rm DM = 15.31845(90)$ cm$^{-3}$ pc \citep{bil16}, which, using the recent model for the Galactic electron density distribution by \citet{yao17}, corresponds to a distance  of 0.89 kpc\footnote{Several previous works adopted a distance of 0.63 kpc, based on \citet{cor02}.}.

\subsection{Previous X-ray results} \label{sec:xr}

X-ray emission from \psrx was discovered with \xmm in 2003,  but the faint flux hampered a detailed analysis \citep{zha05}.
Longer \xmm observations, with simultaneous radio coverage, were carried out in 2011 and led to the  discovery of X-ray variability anti-correlated with the radio intensity \citep{her13}.
These authors found that the X-ray emission was a factor $\sim\!2.5$ {\it brighter}, and pulsed at the star spin period,   during the {\it radio-fainter} Q-mode.
The Q-mode X-ray spectrum was fitted with the  sum of a blackbody with temperature $kT \sim 0.27$ keV and a power law with photon index $\Gamma \sim 2.6$, while either a blackbody or a power law  could fit equally well the spectrum of the fainter B-mode. Using also the timing information in a maximum likelihood analysis, these authors simultaneously derived the spectra of the pulsed and unpulsed emission.  They  found that for the Q-mode the pulsed spectrum could be fit by a blackbody and the unpulsed one by a power law. These components were consistent, respectively, with the blackbody and the power law seen in the two-components fit of the total spectrum.
These results were interpreted assuming that \psrx emits only an unpulsed, non-thermal component in the B-mode, and that the higher luminosity of the Q-mode is caused by the addition of a 100\%-pulsed thermal component \citep{her13}.

The 2011 \xmm data were subsequently reanalyzed by \citet{mer13} and by \citet{sto14}. \citet{mer13} advocated the possibility that thermal X-ray emission is present during both radio modes and suggested that the increased flux of the Q-mode could be due to the appearance of a pulsed non-thermal component. 
They derived a \band{0.6}{1.3} pulsed fraction of $0.56\pm0.08$ in the Q-mode and an upper limit of $0.56$ during the B-mode.
\citet{sto14} showed that the large pulsed fraction of the Q-mode can be explained invoking a magnetized atmosphere on top of the emitting cap (see below).

A further campaign of simultaneous X-ray and radio observations of \psrx was carried out with an \xmm Large Program and the LOFAR, LWA and Arecibo radiotelescopes in November 2014 \citep{mer16}.
Thanks to the larger statistics provided by these data, it was possible to detect X-ray pulsations and to rule out a single power law spectrum also  in the B-mode.
A good fit was obtained either with a single blackbody (with temperature $kT \sim 0.23$ keV) or with a blackbody plus power law, with parameters similar to those of the Q-mode.

\begin{figure}[htbp!]
    \centering
    \includegraphics[width=0.9\linewidth]{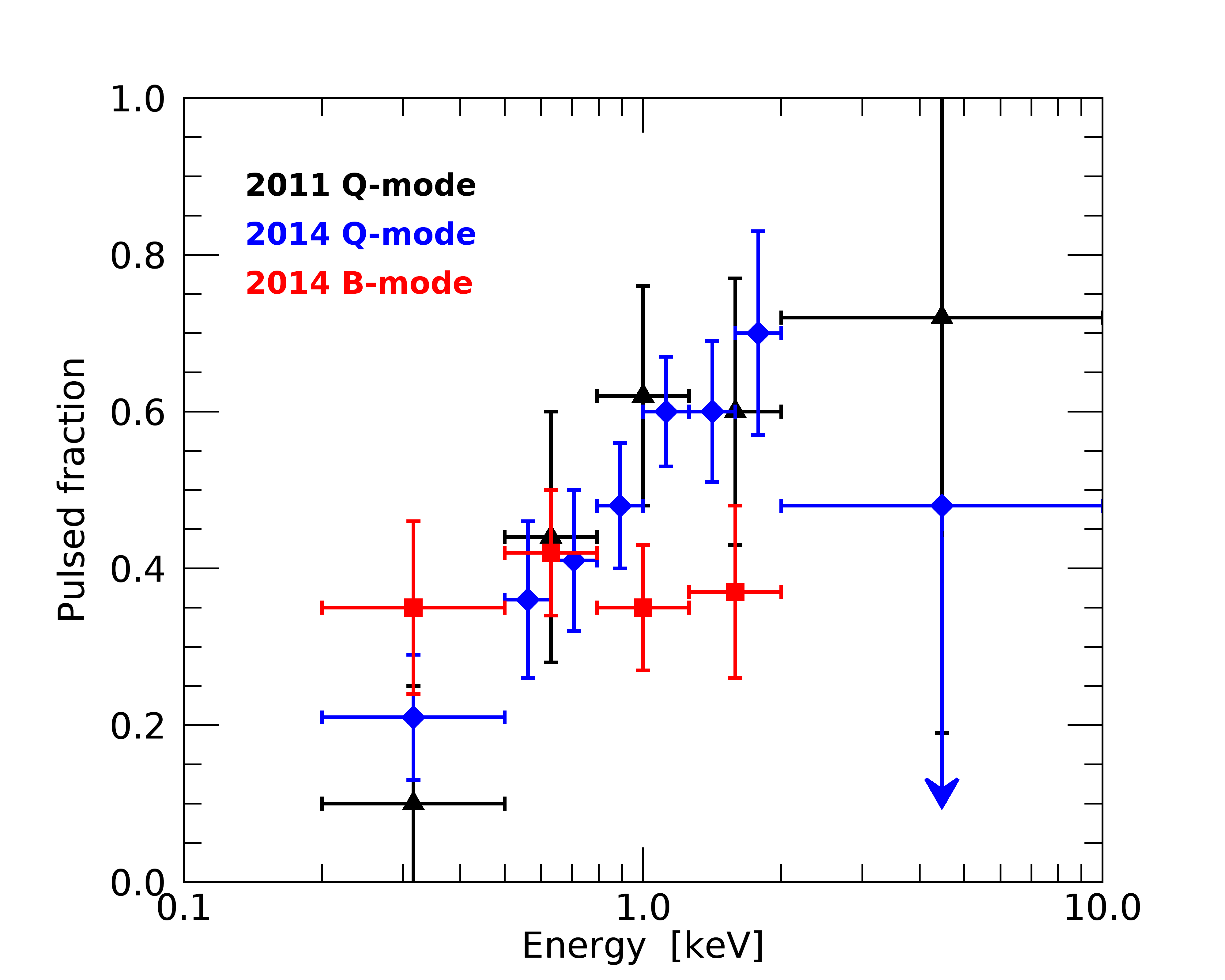}
        \caption{Pulsed fraction of \psrx as a function of energy, as measured by \citet{her13} (Q-mode: black triangles) and by \citet{mer16} (Q-mode: blue diamonds, B-mode: red squares).}
    \label{fig:pf1114}
\end{figure}

The spectral analysis  of the pulsed and unpulsed emission confirmed the findings of \citet{her13}: during the Q-mode, the pulsed emission is thermal, and fitted well by a blackbody, while the unpulsed emission is a power law.
On the other hand, the results were less constraining for what concerns the B-mode:  both  the pulsed and unpulsed emission could be fit by either a  power law or a blackbody.

\citet{her13} and \citet{mer16} derived the pulsed fraction (PF) as a function of energy, defined as

\begin{equation}
	\rm PF(\E) = \frac{c_p(\E)}{c_u(\E)+c_p(\E)},
	\label{eq:pf}
\end{equation}

\noindent
where $\rm c_p$ and $\rm c_u$ are the pulsed and unpulsed counts, respectively, derived with the maximum likelihood method (see below) under the assumption that the pulse shape is sinusoidal.
They found that the PF in the Q-mode increases  steadily from $0.21\pm0.08$ in the \band{0.2}{0.5} band to $0.69\pm0.13$ at 2 keV, while it is nearly constant at $0.38\pm0.05$ between \band{0.2}{2}  in the B-mode (see Figure \ref{fig:pf1114}). If the pulsed flux is due to thermal emission from a polar cap, as indicated by the results of \citet{her13} and \citet{mer16}, such a large PF is difficult to reconcile with the pulsar geometry inferred from the radio data, according to which \psrx is a nearly aligned rotator seen pole-on \citep{des01}, unless magnetic beaming is invoked. Indeed, \citet{sto14} were able to qualitatively reproduce the Q-mode light curve and spectrum, derived from the 2011 observations, using a partially ionized hydrogen atmosphere model with  magnetic field of $2\times10^{12}$ G and  effective temperature of $1.4-1.5$ MK.

\subsection{Geometry from radio data} \label{sec:rad}

The observed properties of pulsars depend on the angles $\xi$ and $\chi$ that the star magnetic axis and the LOS, respectively, make with the spin axis. Introducing the  angle $\eta$ between the magnetic axis and the LOS, and choosing an orthonormal frame with the z axis along the spin axis and the x axis as the projection of the LOS orthogonal to z, it follows that
\begin{equation}
\cos\eta=\cos\xi\cos\chi+\sin\xi\sin\chi\cos\phi\ ,
\end{equation}
where $\phi$ is the rotational phase. When the LOS, the magnetic and rotation axes lie in the same plane,  $\cos\phi=\pm 1$ and  $\eta=\abs{\chi \mp \xi}$. In particular, $\cos\phi=1$
corresponds to the minimum angular distance between the LOS and the magnetic axis. If $\chi < \xi$, this provides $\chi=\xi-\eta$ (inside traverse), while in the opposite case ($\chi > \xi$) it is $\chi=\xi+\eta$ (outside traverse).

The very steep radio spectrum  (S$_{\nu}\propto \nu^{-2.9}$ in the  $0.1-10$ GHz range, \citealt{mal00}) and the presence of drifting subpulses have been considered indications that both  $\xi$  and  $\chi$  are rather small in \psrx . 
A steep spectrum is expected when the LOS  grazes a  beam of radio emission which becomes narrower with increasing frequency.
The drifting subpulses can occur when the LOS crosses nearly tangentially the hollow cone of emission which characterizes the
radio beam of the pulsar. According to the analysis by \citet{des01},
the expected ranges for $\xi$ and $\chi$ are $10^\circ < \xi < 15^\circ$ and $5^\circ < \chi < 10^\circ$.

\setlength{\tabcolsep}{1em}
\begin{table}[htbp!]
\centering \caption{Possible geometries  of \psrx considered in this work}
\label{tab:angles}
\begin{tabular}{lcccc}
\toprule
$\xi$								& 5$^\circ$	& 10$^\circ$	& 20$^\circ$	& 30$^\circ$ 	\\
$\eta \approx \rho_{1\rm\,GHz}$		& 2$^\circ$	& 4$^\circ$		& 6$^\circ$ 	& 8$^\circ$ 	\\
$\chi = \xi-\eta$ (inside traverse)	& 3$^\circ$	& 6$^\circ$ 	& 14$^\circ$ 	& 22$^\circ$ 	\\
$\chi = \xi+\eta$ (outside traverse)& 7$^\circ$	& 14$^\circ$	& 26$^\circ$ 	& 38$^\circ$ 	\\
\bottomrule
\end{tabular}

$\xi$ is the angle between the magnetic and spin axis;

$\chi$ is the angle between the line of sight  and the spin axis.
\end{table}

Recently, \citet{bil18} reconsidered the problem of the geometry of \psrx. The angles $\xi$ and $\chi$ are evaluated by analyzing the frequency dependence of the observed profile width $w_\nu$ in the radio band. This is related to the frequency-dependent opening angle of the emission cone $\rho_\nu$ via:
\begin{equation}
 \cos\rho_\nu = \cos\xi \cos(\xi+\eta) + \sin\xi \sin(\xi+\eta) \cos(w_\nu/2)
\end{equation}
 \noindent
 \citep{gil84,ran93}. $\rho_\nu$ depends on the frequency as
\begin{equation}
\rho_\nu = \rho_\infty (1+K\nu_{\rm GHz}^{-a}) P^{-1/2},
\end{equation} where $K=0.066 \pm 0.010$, $a=1.0 \pm 0.1$ and $\rho_\infty$ is the opening angle of the emission cone at infinite radio frequency when $P=1$~s \citep{mit99}.

\citet{des01} considered $\rho_{1\rm\,GHz}=\rho_\infty\times 1.066\times P^{-1/2}$ in a range between $4^\circ.3$ and $5^\circ.7$ and excluded the outside traverse case because, in the context of the rotating carousel model adopted by these authors, it made inconsistent predictions on the periodic modulation of drifting subpulses.
 
On the other hand, \citet{bil18} noted that the uncertainties on $\rho_{1\rm\,GHz}$  are larger and, following \citet{mit99}, adopted $2^\circ < \rho_{1\rm\,GHz} < 8^\circ$. She found that $\eta \approx \rho_{1\rm\,GHz}$ and considered that both branches of solutions corresponding to the inside and the outside traverse should be retained because 
the amplitude modulation at $37P$ of the drifting subpulses was not detected in her new radio data.
 
In conclusion, the angle between magnetic and spin axis derived by \citet{bil18} is constrained in the range: $5^\circ < \xi < 30^\circ$. For each  value of $\xi$ only two well defined values of $\chi$ are possible and are in the ranges $3^\circ < \chi < 22^\circ$ (inside traverse model) or  $7^\circ < \chi < 38^\circ$ (outside traverse model). In the following, we consider four representative pairs of $\xi$ and $\chi$ for the inside traverse case and the corresponding ones for the outside traverse, as indicated in Table \ref{tab:angles}.

\section{The ray-tracer code} \label{sec:model}

Our computation of  the phase-dependent spectrum emitted by the magnetic polar caps of a NS, as seen by a distant observer, is 
done in two steps: the first involves the computation of the local spectrum emitted by each patch of the surface, while the second requires the collection of the contributions of surface elements which are in view at different rotation phases.
Details of each step are presented in \citet[ see also \citealt{tav15}]{zan06} and the numerical calculation was carried out using an \textsc{idl} script; here we discuss some specific assumptions for the \psrx case.

While blackbody emission is isotropic and depends only on the surface temperature $T_s$, the spectrum emerging from a magnetic atmosphere or a solid condensed surface depends on the magnetic field  intensity and orientation with respect to the local normal ($\theta_B$), as well as on the angles that the photon direction makes with the surface normal ($\theta_k$ and $\phi_k$). The emerging radiation is therefore beamed.
However, if we consider magnetic polar caps that are sufficiently small, they can be treated as point-like and therefore the normal to the emitting surface is aligned with the dipolar magnetic field ($\theta_B=0^\circ$), implying also azimuthal symmetry (no dependence on $\phi_k$). The point-like assumption is justified if the semi-aperture of the cap is $\lesssim\!5^\circ$ (\citealt{tur13}, see also \citealt{bel02}).
Indeed, the observed radius of the emitting region, as derived from spectral fits, is in the range $30-300$ m (see sections \ref{sec:atmo} and \ref{sec:solid}), which corresponds to a semi-aperture $\sim 0^\circ.15 - 1^\circ.5$, small enough to treat the caps as point-like.

Once the emission model is specified, the spectrum at infinity is computed collecting the contributions from each patch, accounting for general relativistic effects. Because of ray bending, the visible part of the star surface is larger than one hemisphere and, for certain geometrical configurations, both caps are fully in view at the same phase (e.g. \citealt{pec83,bel02,tur13}).

We give all the spectral results in terms of observed quantities at infinity. The observed temperature, $T_{\infty}$, is related to that measured by a stationary observer at the star surface, $T_{s}$, by $T_{\infty} = T_s\sqrt{1 - 2GM/R c^2}$, while the observed polar cap radius is inferred from the equations of \citet{tur13}.  We assume a neutron star of mass $M = 1.5 \msun$ and radius $R = 12$ km, and adopt for \psrx a distance of 0.89 kpc.

\section{Observations and data analysis}
 \label{sec:obs}

\setlength{\tabcolsep}{1.5em}
\begin{table*}[htbp!]
\centering \caption{Exposure times and number of detected counts for \psrx in the different radio modes}
\label{tab:obs}
\begin{tabular}{lcccc}
\toprule
Year        & Radio mode    & Epic camera	& Exposure time & Counts$~^{\rm a}$           \\
            &               &               & ks            & \band{0.2}{10}    \\
\midrule
2011		&	Q			& pn			& 48.5			& $590 \pm 40$	\\
2011		&	Q			& MOS			& 53.4			& $293 \pm 26$	\\
2011		&	B			& pn			& 40.9			& $191 \pm 26$	\\
2011		&	B			& MOS			& 45.2			& $99 \pm 17$	\\
\midrule
2014		&	Q			& pn			& 123.6			& $1450 \pm 66$	\\
2014		&	Q			& MOS			& 133.8			& $680 \pm 40$	\\
2014		&	B			& pn			& 174.2			& $944 \pm 61$	\\
2014		&	B			& MOS			& 189.9			& $410 \pm 35$	\\
\midrule
2011+2014   &   Q          	& pn            & 172.5         & $2054 \pm 77$ \\
2011+2014   &   Q         	& MOS			& 187.2         & $973 \pm 48$  \\
2011+2014   &   B          	& pn            & 215.6         & $1134 \pm 66$ \\
2011+2014   &   B         	& MOS			& 235.1         & $512 \pm 39$  \\
\bottomrule
\end{tabular}

$^{\rm a}$ Total (background-subtracted) source counts derived with the maximum likelihood method.

\end{table*}

\xmm observed \psrx in 2003, 2011 and 2014. In all the analyses reported here, we used the sum of the 2011 and 2014 observations, after checking that the individual data sets give consistent results. 
From the exposure times and number of source counts reported in Table~\ref{tab:obs} one can see the increase in the statistics, compared to previous works based on the individual data sets (more details on the 2011 and 2014 observations can be found in \citealt{her13} and \citealt{mer16}, respectively).

Although there are no simultaneous radio data for the  2003 observations, the average X-ray flux measured in   December 2003  suggests that the pulsar was in the B-mode for most of the time \citep{mer17}. In principle, 
we could have added these data to those analyzed here, but also in view of  their short exposure, we decided to restrict our analysis to the 2011 and 2014 data sets for which the mode identification based on radio observations is certain.

During all observations, the EPIC pn camera was operated in full frame mode, which provides a time resolution of 73 ms, while  the
two MOS cameras were used in small window mode (the imaging mode with the highest time resolution, 0.3 s). For the three cameras
the thin optical filter was used.

We reprocessed the EPIC pn data using the SAS task {\it epreject} to reduce the detector noise at the lowest energies. To remove the
periods of high background we used the same cuts adopted in \citet{mer16}, i.e.  we excluded all the time intervals with a pn
count rate in the range  \band{10}{12}  higher than 1.2 cts~s$^{-1}$. The resulting net exposure times are given in
Table~\ref{tab:obs}.
To separate the data of the B- and Q-modes, we used the times derived from radio data in the previous works by \citet{her13} and \citet{mer16}.

We used single- and multiple-pixel events for both the pn and MOS. The events detected in the two MOS cameras were combined into a single data set, and analyzed with averaged exposure maps and response files.

To extract the source counts and spectra, we used a maximum likelihood (ML) technique, as first introduced for this pulsar by \citet{her13}.
Briefly, this consists in estimating the most probable number of source and background counts by exploiting the knowledge of the
instrumental point-spread function (PSF), derived from in-flight
calibrations\footnote{http://www.cosmos.esa.int/web/xmm-newton/calibration/documenta-tion},
suited for the average energy in the considered bin. This method has the advantage of maximizing the number of source counts and of using a local background estimated at the effective  position of the source.

We applied the ML analysis to a circular region centered at \coord{09}{46}{07}{8}{+09}{52}{00}{8}, with a radius of $30''$, from which we excluded a circle of radius $30''$  centered at \coord{09}{46}{10}{7}{+09}{52}{26}{4} to avoid a nearby source.
The energy bins for the ML analysis  were chosen in such a way to have a well determined background\footnote{We required a ratio between the background value and its error $>5$.} and a source significance greater than $5\sigma$\footnote{Except for  the highest
energy bin,  where it was not possible to reach $5\sigma$,  and for which we  adopted  the  upper boundary yielding  the largest
source significance.} in each spectral channel. 
In this way we obtained 23 energy bins for the pn (\band{0.2}{7}) and 8 for the MOS (\band{0.2}{4.3}) for the Q-mode, and
13 energy bins for the pn (\band{0.2}{6.7}) and 7 for the MOS (\band{0.2}{5.5}) for the B-mode. 

In the spectral fits we used the interstellar absorption model \textsc{phabs} of XSPEC, version 12.8.2.
All the errors are  at 1$\sigma$ level.
For the timing analysis, the pulse phases of \psrx counts were computed using the ephemeris
given in \citet{mer16}, that are valid from 54861.014 to 57011.249
MJD (Jan 2009 to Dec 2014). 

\bigskip

\begin{figure*}[htbp!]
    \centering
    \includegraphics[width=0.8\linewidth]{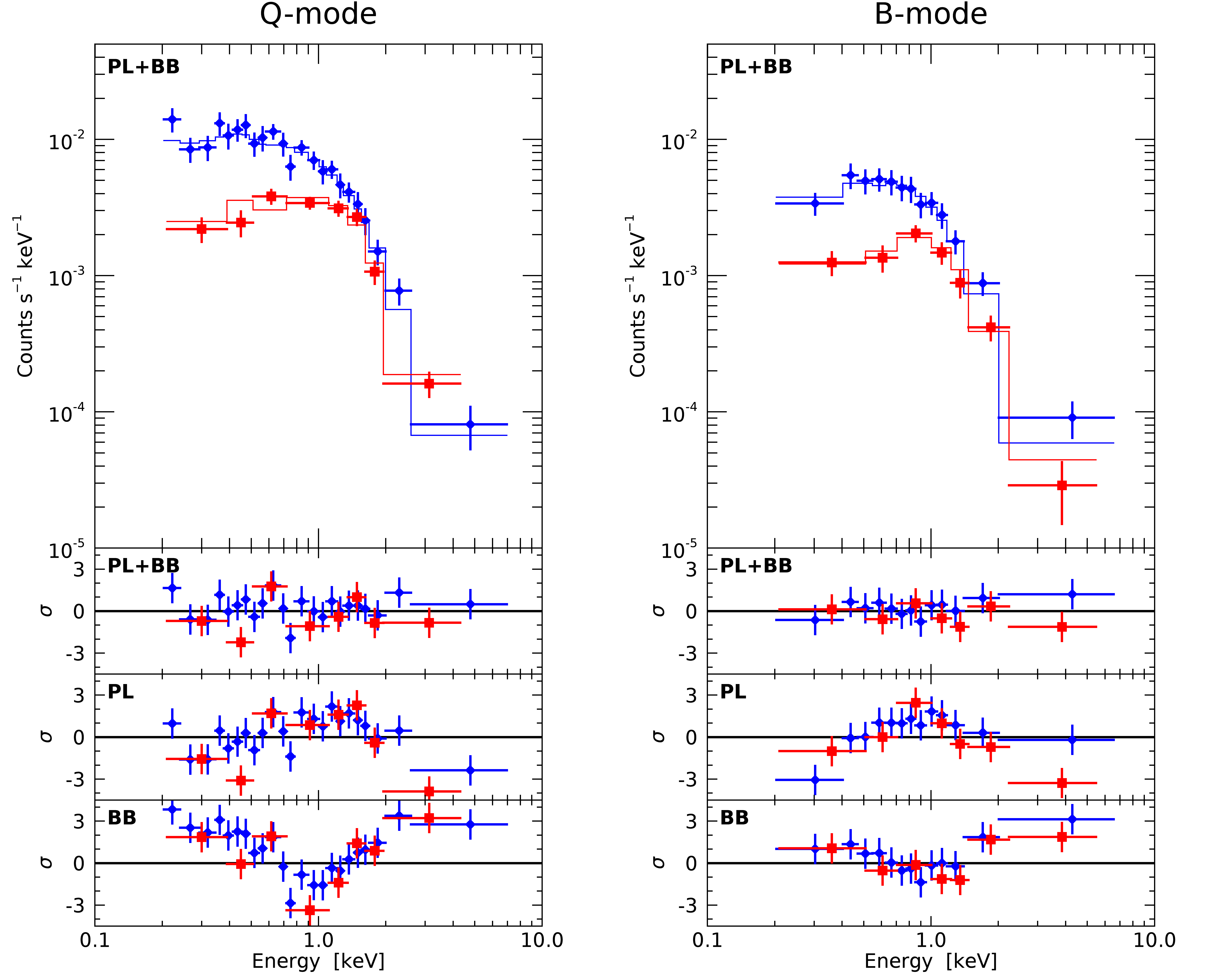}
    \caption{EPIC-pn (blue diamonds) and -MOS (red squares) X-ray phase-averaged spectra of \psrx in the Q-mode (left) and in the B-mode (right). The top panels show the best fit using absorbed power law plus blackbody models; the lower panels show the residuals of the best fit (PL+BB), of an absorbed power law model (PL) and of an absorbed blackbody model (BB) in units of $\sigma$.}
    \label{fig:ressingle}
\end{figure*}

 \label{sec:bb}

\section{Results}
\subsection{Blackbody thermal emission}

As a first step,  we considered fits to the total (i.e. pulsed plus unpulsed) emission of \psrx  using only power law and blackbody spectral components. 
We obtained best fit parameters (Table \ref{tab:ML2D}) fully consistent with those found in previous analyses \citep{her13,mer16} and in general with slightly smaller uncertainties, thanks to the better statistics provided by joining the  2011 and 2014 data.

In the Q-mode,  single-component models are clearly rejected (see Figure \ref{fig:ressingle}, left panel), while good fits are
obtained with a blackbody plus power law or with the sum of two blackbody components. In the B-mode, a  single power law is clearly rejected ($\chi_\nu^2 = 2.28/18$ degrees of freedom, dof, corresponding to a  null hypothesis probability nhp
= 0.004). A blackbody model with temperature $kT = 0.22 \pm 0.01$
keV is marginally acceptable ($\chi_\nu^2 = 1.66/18$ dof, nhp =
0.04),  but the shape of the residuals shown in  the right panels
of Figure \ref{fig:ressingle} indicates that a second
spectral component is needed. In fact, similarly to the Q-mode,  a good fit
can be obtained by using either a power law plus a blackbody or
the sum of two blackbodies (see Table \ref{tab:ML2D}).

\setlength{\tabcolsep}{2em}
\begin{table*}[htbp!]
\centering \caption{Best fit parameters for the phase-averaged spectra of the  Q- and B- modes}
\label{tab:ML2D}

\begin{tabular}{lcccc}
\toprule
                            & Q-mode                & Q-mode                & B-mode                & B-mode                \\[5pt]
                            & PL + BB               & BB + BB               & PL + BB               & BB + BB               \\[5pt]
\midrule\\[-5pt]
$\Gamma$                    & $2.6_{-0.1}^{+0.2}$   & \dots                 & $2.2_{-0.3}^{+0.2}$   & \dots                 \\[5pt]
$K~{}^{\rm a}$              & $2.9_{-0.5}^{+0.4}$   & \dots                 & $1.0 \pm 0.3$         & \dots                 \\[5pt]
$\rm Flux_{PL}^{0.5-2}$     & $6.6_{-1.1}^{+0.9}$   & \dots                 & $2.3_{-0.8}^{+0.6}$   & \dots                 \\[5pt]
$\rm Flux_{PL}^{0.2-10}$    & $18 \pm 2$            & \dots                 & $6 \pm 1$             & \dots                 \\[5pt]
\midrule\\[-5pt]

$kT_1$ (keV)                & $0.30 \pm 0.02$       & $0.35_{-0.02}^{+0.03}$& $0.21 \pm 0.02$       & $0.48 \pm 0.08$       \\[5pt]
$R_{\rm BB_1}~{}^{\rm b}$ (m)   & $27_{-4}^{+5}$        & $25 \pm 4$            & $41_{-9}^{+10}$       & $6_{-2}^{+3}$         \\[5pt]
$\rm Flux_{BB_1}^{0.5-2}$   & $6.0 \pm 1.1$         & $9.2_{-1.2}^{+0.9}$   & $3.4_{-0.8}^{+0.7}$   & $1.7_{-0.7}^{+0.6}$   \\[5pt]
$\rm Flux_{BB_1}^{0.2-10}$  & $7.6 \pm 1.4$         & $12 \pm 1$            & $4 \pm 1$             & $3.2_{-0.7}^{+0.6}$   \\[5pt]
\midrule\\[-5pt]

$kT_2$ (keV)                & \dots                 & $0.12 \pm 0.01$       & \dots                 & $0.17 \pm 0.01$       \\[5pt]
$R_{\rm BB_2}~{}^{\rm b}$ (m)   & \dots                 & $200_{-40}^{+55}$     & \dots                 & $72_{-9}^{+11}$       \\[5pt]
$\rm Flux_{BB_2}^{0.5-2}$   & \dots                 & $3.75_{-1.25}^{+1.05}$& \dots                 & $4.1_{-0.7}^{+0.5}$   \\[5pt]
$\rm Flux_{BB_2}^{0.2-10}$  & \dots                 & $9.2 \pm 1.0$         & \dots                 & $6.2_{-0.7}^{+0.5}$   \\[5pt]
\midrule\\[-5pt]

$\rm Flux_{TOT}^{0.5-2}$    & $12.6 \pm 0.4$        & $13.0 \pm 0.4$        & $5.7 \pm 0.3$         & $5.85 \pm 0.25$       \\[5pt]
$\rm Flux_{TOT}^{0.2-10}$   & $25.9 \pm 1.1$        & $21.3 \pm 0.8$        & $10.8 \pm 0.8$        & $9.4 \pm 0.5$         \\[5pt]

$\chi_{\nu}^2$/dof          & 1.08/27               & 1.25/27               & 0.51/16               & 0.43/16               \\[5pt]
nhp                      & 0.36                  & 0.18                  & 0.94                  & 0.98                  \\[5pt]
\bottomrule
\end{tabular}

\raggedright
Joint fits of pn + MOS spectra with \nh fixed to $4.3 \times 10^{20}$ cm$^{-2}$. PL = power law, BB = blackbody. The fluxes, corrected for the absorption, are  in units of $10^{-15}$~erg~cm$^{-2}$~s$^{-1}$. Errors at $1\sigma$.\\
$^{\rm a}$ Normalization of the power law at 1 keV in units of $10^{-6}$~photons~cm$^{-2}$~s$^{-1}$~keV$^{-1}$.\\
$^{\rm b}$ Blackbody radius for an assumed distance of 0.89 kpc.\\
\end{table*}

\begin{figure}[htbp!]
    \centering
    \includegraphics[width=0.9\linewidth]{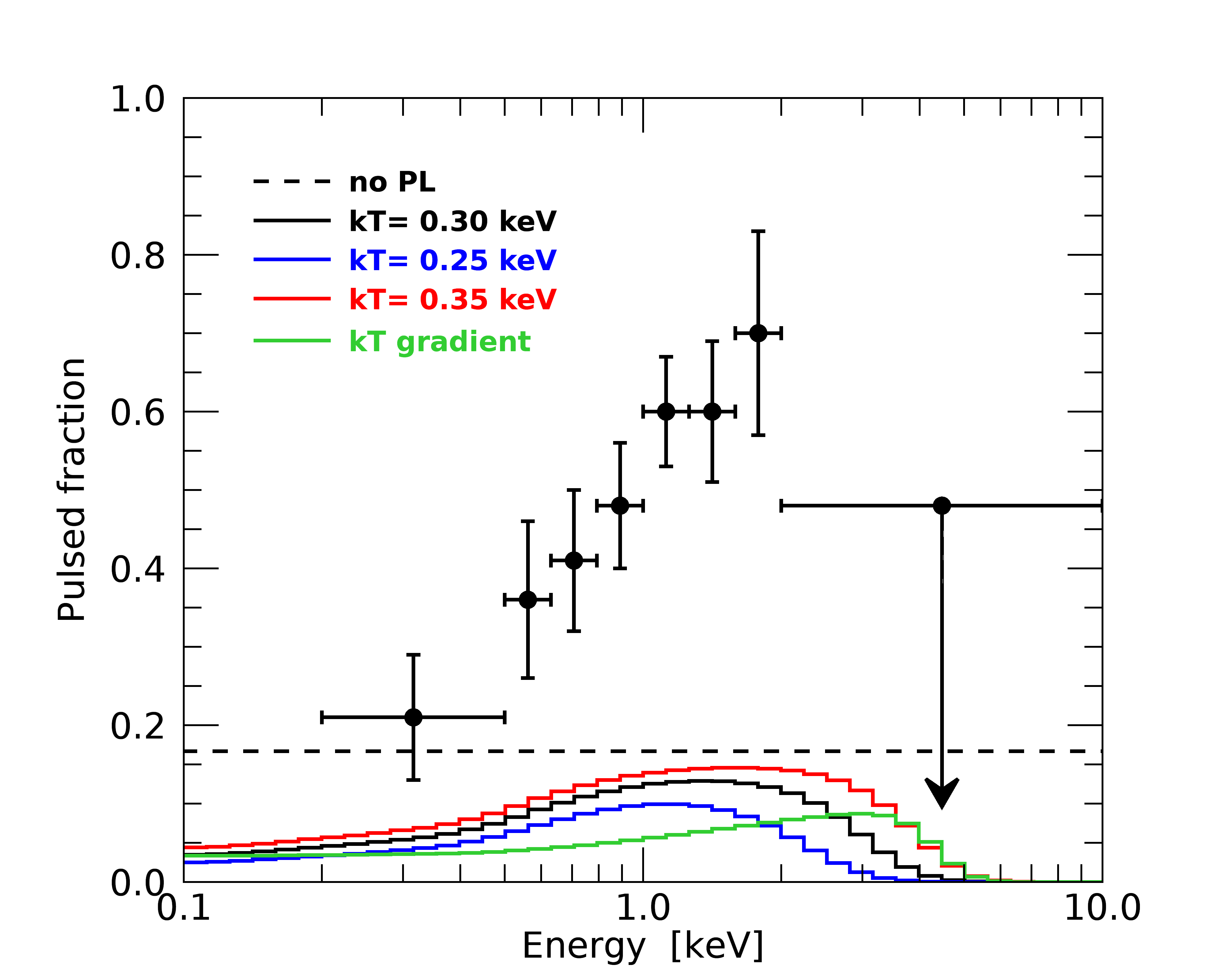}
    \caption{Pulsed fraction as a function of energy computed for the case of blackbody thermal emission and the geometrical configuration of Table~\ref{tab:angles} producing the highest modulation ($\xi=30^\circ$, $\chi=38^\circ$). The dashed line is the PF expected if only the blackbody emission from the polar cap is present, while the solid lines show the PF given by a polar cap at temperatures $kT=0.30$ keV (black), $kT=0.25$ keV (blue), $kT=0.35$ keV (red) when also a non-thermal unpulsed emission is present (power law with $\Gamma =2.6$). The pulsed fraction produced by the non-uniform temperature distribution of eq. \ref{eq:ts} is shown by the green line. The black dots indicate the observed  pulsed fraction of the Q-mode.}
    \label{fig:pfBB}
\end{figure}

\setlength{\tabcolsep}{2em}
\begin{table*}[htbp!]
\centering \caption{Best fit parameters for the spectra of the pulsed and unpulsed emission in the two radio modes}
\label{tab:ML3D}
\begin{tabular}{lcccc}
\toprule
                            & Q-mode                & Q-mode            & B-mode                & B-mode                \\[5pt]
                            & Unpulsed              & Pulsed            & Unpulsed              & Pulsed                \\[5pt]
                            & PL                    & BB                & PL                    & BB                    \\[5pt]
\midrule\\[-5pt]
$\Gamma$                    & $2.50 \pm 0.15$       & \dots             & $2.3 \pm 0.2$			& \dots                 \\[5pt]
$K~^{\rm a}$                & $3.10 \pm 0.25$       & \dots             & $1.7 \pm 0.2$			& \dots                 \\[5pt]

\midrule\\[-5pt]
$kT$ (keV)                  & \dots                 & $0.27 \pm 0.02$   & \dots                 & $0.23_{-0.03}^{+0.05}$\\[5pt]
$R_{\rm BB}~^{\rm b}$ (m)   & \dots                 & $32_{-5}^{+6}$    & \dots                 & $25_{-8}^{+10}$		\\[5pt]

\midrule\\[-5pt]
$\rm Flux^{0.5-2}$          & $7.1 \pm 0.5$			& $5.7 \pm 0.5$		& $3.8 \pm 0.4$		    & $1.9 \pm 0.3$			\\[5pt]
$\rm Flux^{0.2-10}$         & $19 \pm 1$            & $7.1 \pm 0.6$		& $10 \pm 1$			& $2.4 \pm 0.4$			\\[5pt]
$\chi_{\nu}^2$/dof          & 0.797/6               & 0.691/6           & 0.266/5               & 0.200/5               \\[5pt]
nhp                      & 0.57                  & 0.66              & 0.93                  & 0.96                  \\[5pt]
\bottomrule
\end{tabular}

\raggedright
Joint fits of pn + MOS spectra with \nh fixed to $4.3 \times 10^{20}$ cm$^{-2}$. PL = power law, BB = blackbody. The fluxes, corrected for the absorption, are  in units of $10^{-15}$~erg~cm$^{-2}$~s$^{-1}$. Errors at $1\sigma$.\\
$^{\rm a}$ Normalization of the power law at 1 keV in units of $10^{-6}$~photons~cm$^{-2}$~s$^{-1}$~keV$^{-1}$.\\
$^{\rm b}$ Blackbody radius for an assumed distance of 0.89 kpc.\\
\end{table*}

We also examined the spectra of the pulsed and unpulsed emission in the two radio modes, using a ML analysis that also takes into account the timing information of each photon (see details in \citealt{her17} and \citealt{rig18}). The results,
again in agreement with those obtained with the 2014 data alone, are summarized in Table \ref{tab:ML3D}.

Although modeling the thermal emission with blackbody components
gives formally acceptable results from  the point of view of the
spectral fits,  there are obvious problems to reproduce the
observed energy-dependence of the folded pulse profiles. In fact,
the blackbody emission from an element of the NS surface is
isotropic and the light curves produced by a rotating hot spot do
not depend on energy. Therefore, the pulsed fraction depends only on the geometrical parameters and on the compactness ratio $M/R$ of the star.

Energy-dependent pulse profiles can be obtained if an unpulsed power law component is added to the blackbody, as shown in Figure \ref{fig:pfBB} for emitting polar caps of different temperatures.  
The figure refers to the geometric configuration of Table~\ref{tab:angles}  yielding the highest modulation, i.e. $\xi=30^\circ$ and $\chi=38^\circ$.
In all cases, we adopted a polar cap size consistent with the spectral results of the Q-mode and included the corresponding best-fit unpulsed power law.   
The dashed line gives for comparison the pulsed fraction that would be obtained  in the absence of the unpulsed power law component.

For completeness, even if not of direct interest for the case of \psrx, Figure \ref{fig:pfBB} shows also the pulsed  fraction expected in the case of
thermal emission from the whole NS surface with an inhomogeneous temperature $T_s$ given by
\begin{equation}
    T_s = T_p \abs{\cos \theta}^{1/2},
    \label{eq:ts}
\end{equation}

\noindent where $T_p$ is the temperature at the magnetic poles and $\theta$ is the magnetic colatitude.
Equation \ref{eq:ts} follows from the more general result of \citet{gre83}, by assuming a dipolar magnetic field, a locally plane-parallel geometry of the heat-insulating layer, and a negligibly small thermal conductivity perpendicular to the magnetic field.
As expected, such a large emitting area produces an even smaller pulsed fraction.\\

\subsection{Magnetized hydrogen atmosphere} \label{sec:atmo}

The surface of a neutron star can be covered by a gaseous atmosphere, where radiative processes alter the emergent spectrum.
Model atmospheres of highly magnetized neutron stars were computed by many authors, following the pioneering work of \citet{shi92} (see, e.g., \citealt{pot16} for a review and references therein).
Here we use the approach of \citet{sul09} and corresponding code to compute a grid of model atmospheres in the required ranges of surface effective temperature $T_\mathrm{eff}$ and surface magnetic field. In a magnetized plasma, the electromagnetic radiation propagates in the form of extraordinary (X) and ordinary (O) normal modes, that have different opacities and polarization vectors \citep{gin70,mes92}. The models are computed in the plane parallel approximation.

We assume that the atmosphere consists of hydrogen, and calculate the equation of state of partially ionized hydrogen plasma and its polarization-dependent opacities according to \citet{pot03} with improvements introduced by \citet{pot14}.
The main improvement consists in the inclusion of radiative transitions from excited bound states of magnetized H atoms. The plasma polarizabilities and normal-mode opacities are computed using the Kramers-Kronig relation, as in \citet{pot04}. We take into account partial mode conversion due to vacuum polarization as described by \citet{van06}.

The radiation transfer equation was solved for about 200 photon energies from 0.01 to 40 keV at 40 angles to the atmosphere normal, uniformly distributed on a logarithmic scale from $1^\circ$ to $89^\circ.9$ with the addition of a further point at $0^\circ.1$.  Altogether 84 model atmospheres were computed for seven surface effective temperatures ({0.5, 0.8, 1, 1.2, 1.5, 2 and 3 MK) and twelve values of the magnetic field (1, 1.5, 2, 2.4, 2.7, 3, 3.5, 4, 5, 6, 7 and 8 $\times 10^{12}$ G). The surface gravity $\log g = 14.241$ was fixed to the adopted NS parameters, $M = 1.5\msun$ and $R=12$ km. The magnetic field  was assumed to be normal to the surface.
As argued in section~\ref{sec:model}, this is a good approximation for small hot spots located around the magnetic poles.

\begin{figure}[htbp!]
    \centering
    \includegraphics[width=0.9\linewidth]{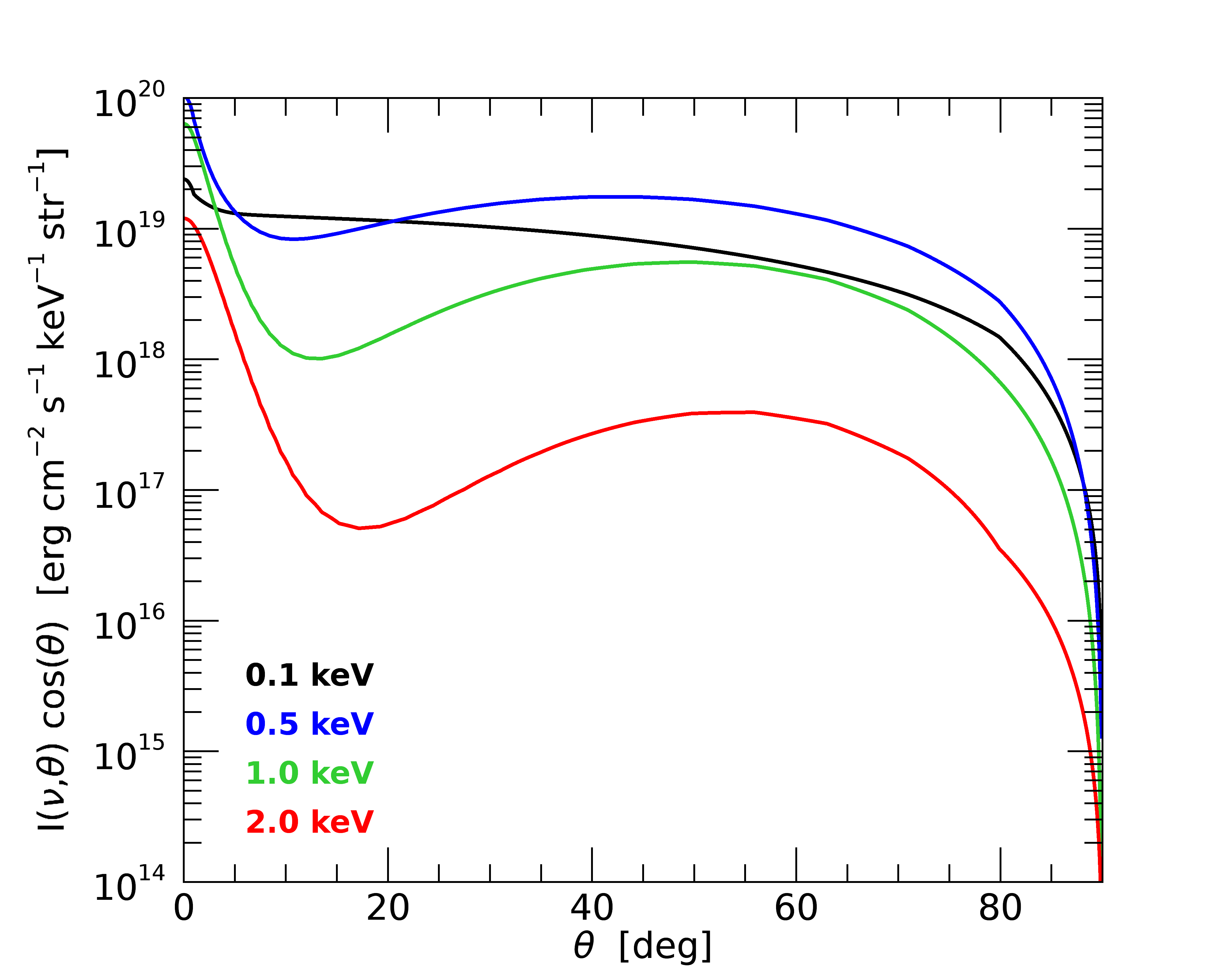}
    \caption{Angular dependence of the emerging intensity for a hot spot covered by a magnetized hydrogen atmosphere with $T_\mathrm{eff} = 1$~MK and $B = 4\times10^{12}$~G. The curves refer to different photon energies.}
    \label{fig:beam}
\end{figure}

Figure~\ref{fig:beam} shows the angular dependence of the specific intensity at the magnetic pole for different photon energies in the case of $T_\mathrm{eff} = 1$~MK and $B= 4\times10^{12}$~G.
The angular dependence we find is typical of magnetized atmospheres and very similar to those obtained previously, e.g., by \citet{pav94} and \citet{zav02}.

The pronounced peak along the magnetic field is related to the reduction in the opacity of both polarization modes at small angles between the magnetic field and the direction of photon propagation \citep{sul09}. However, the total flux in that peak is relatively small because it occupies a small solid angle. Most of the radiation escapes in the second broad maximum at intermediate angles, that gives rise to the so-called fan-beamed emission. This off-axis maximum is caused by the opacity reduction in the X-mode at large angles \citep{pav94,sul09}. Finally, we note a gradual softening of the specific intensity at higher angles  approaching 90$^\circ$, because relatively cold surface atmospheric layers contribute substantially at these angles. The emerging specific intensity becomes more anisotropic with increasing photon energies, as the relative importance of electron scattering to the total opacity increases. 
 
For each set of parameters explored in our model, we computed the phase-averaged spectrum and implemented it in XSPEC. We found that, for any of the considered geometries (Table~\ref{tab:angles}), it was possible to find an acceptable fit to the phase-averaged spectra of both the Q- and B-mode using only the atmosphere model, without the need of an additional power law component. 

The best fit parameters depend on  the $\xi$ and $\chi$ values. 
For all the considered angles, the fits to the Q-mode spectra gave a small absorption, consistent with 0 ($1\sigma$ upper limit of \nh$<6 \times 10^{19}$ cm$^{-2}$). The best fit spectral parameters were in the ranges $B = 2 - 6 \times 10^{12}$ G, $kT = 0.10 - 0.15$ keV, and  $R_{\rm cap} = 150 - 300$ m. In the B-mode the absorption was poorly constrained; therefore we fixed it to $6 \times 10^{19}$ cm$^{-2}$ and obtained best fit values of $kT$ similar to those of the Q-mode, but with $R_{\rm cap}$ in the range $100 - 200$ m and unconstrained values of $B$. 
For the Q-mode the best fit was found for $\xi=5^\circ$ and $\chi=3^\circ$, while all the considered geometries gave equally good fits ($\chi_\nu^2 \approx 1$) to the B-mode spectra.

In order to distinguish between the different possibilities allowed by the atmosphere fits to the phase-averaged spectra, we examined the pulsed profiles, which have a stronger dependence on the geometrical configuration. For each pair of $\xi$ and $\chi$ values we computed the expected pulse profile in the energy range \band{0.5}{2}, where the pulsation is detected with the highest significance, taking into account the instrumental response of EPIC. For each geometrical configuration and mode we used the corresponding best fit values of $kT$, $R_{\rm cap}$ and $B$ derived in the spectral analysis.  
The resulting pulse profiles were quantitatively compared to the observed ones using a Kolmogorov-Smirnov (KS) test. In this way, we found that the acceptable configurations  (probability $>10\%$ that the observed data come from the model) are $\xi=5^\circ$, $\chi=3^\circ$ and $\xi=5^\circ$, $\chi=7^\circ$ for the Q-mode. The B-mode analysis adds no information because all the configurations were acceptable (KS-test probability $>13 \%$).

More information can be obtained by performing phase-resolved spectroscopy. For this, we extracted with the ML technique the Q- and B-mode spectra of \psrx in two phase intervals of duration $0.5$ centered at phase $0$ (pulse maximum)  and at phase $0.5$ (pulse minimum). 
To fit these spectra, we used  models specifically computed by integrating the predicted emission over the corresponding phase-intervals as described above.

The joint fit of the four Q-mode spectra (two pn and two MOS) showed that the geometrical configurations with small angles are preferred (nhp = 0.06 for $\xi=5^\circ$ and $\chi=3^\circ$, nhp $<3\times10^{-3}$ for all the other cases), although the fit is worse than that of the phase-averaged spectra.

Better fits could be obtained by adding to the model a power law component, that was assumed to be unpulsed by linking its parameters to common values in the two phase bins.
We initially let the interstellar absorption as a  free parameter, but since it was poorly constrained, we finally  fixed it to the value of $4.3 \times 10^{20}$ cm$^{-2}$ used in previous analysis. 
In the Q-mode, for all the considered geometrical configurations, good fits were obtained with a photon index $\Gamma \approx 2-2.5$. 
However, for $\xi \geq 20^\circ$ the best-fit models required values of magnetic field smaller than $10^{12}$ G, inconsistent with the value expected from the timing parameters of \psrx. All the expected pulse profiles, together with the spectral parameters from which they are computed and the KS-test probability, are shown in Figure \ref{fig:lc_50}.

\begin{figure*}[htbp!]
    \centering
    \includegraphics[height=9cm]{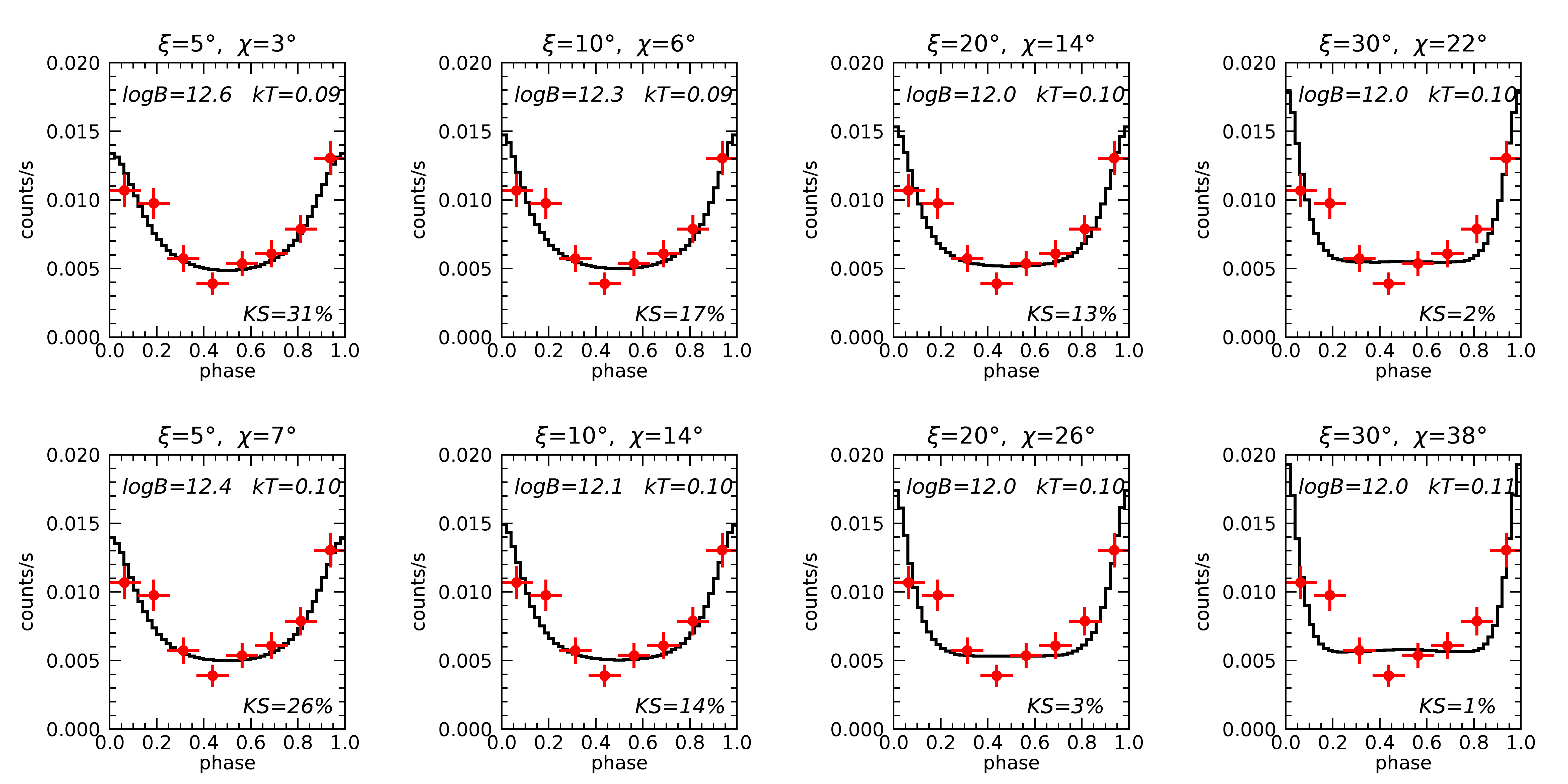}
    \caption{Expected pulse profiles in the \band{0.5}{2} range (solid lines), in the case of a hydrogen atmosphere model, for different pairs of $\xi$ and $\chi$ and the corresponding best fit spectral parameters for the Q-mode spectra. The KS-test probability is also shown. Upper panels: inside traverse, lower panels: outside traverse. The red dots with error bars show the observed Q-mode data.}
    \label{fig:lc_50}
\end{figure*}

The configuration favoured by both the KS test for the light curves and the $\chi^2$ test for the spectra is $\xi=5^\circ$ and $\chi=3^\circ$, that yields $\Gamma=2.5\pm0.2$, $kT=0.089_{-0.005}^{+0.014}$ keV, $R_{\rm cap}=260_{-70}^{+60}$ m and $B=(4.0_{-0.7}^{+0.9}) \times 10^{12}$ G (see Table~\ref{tab:model} for all the details). Figure \ref{fig:specAT} shows the best-fit phase-averaged (left panel) and phase-resolved (right panel) spectra. As a comparison, in the lower panels the residuals of the atmosphere model alone are shown: while in the phase-averaged case the addition of the power law is not required, the fit of the phase-resolved spectra (especially the spectra of the pulse maximum) is significantly improved.

\begin{figure*}[htbp!]
    \centering
    \includegraphics[trim=0cm 2.5cm 0cm 0cm,clip,width=0.8\linewidth]{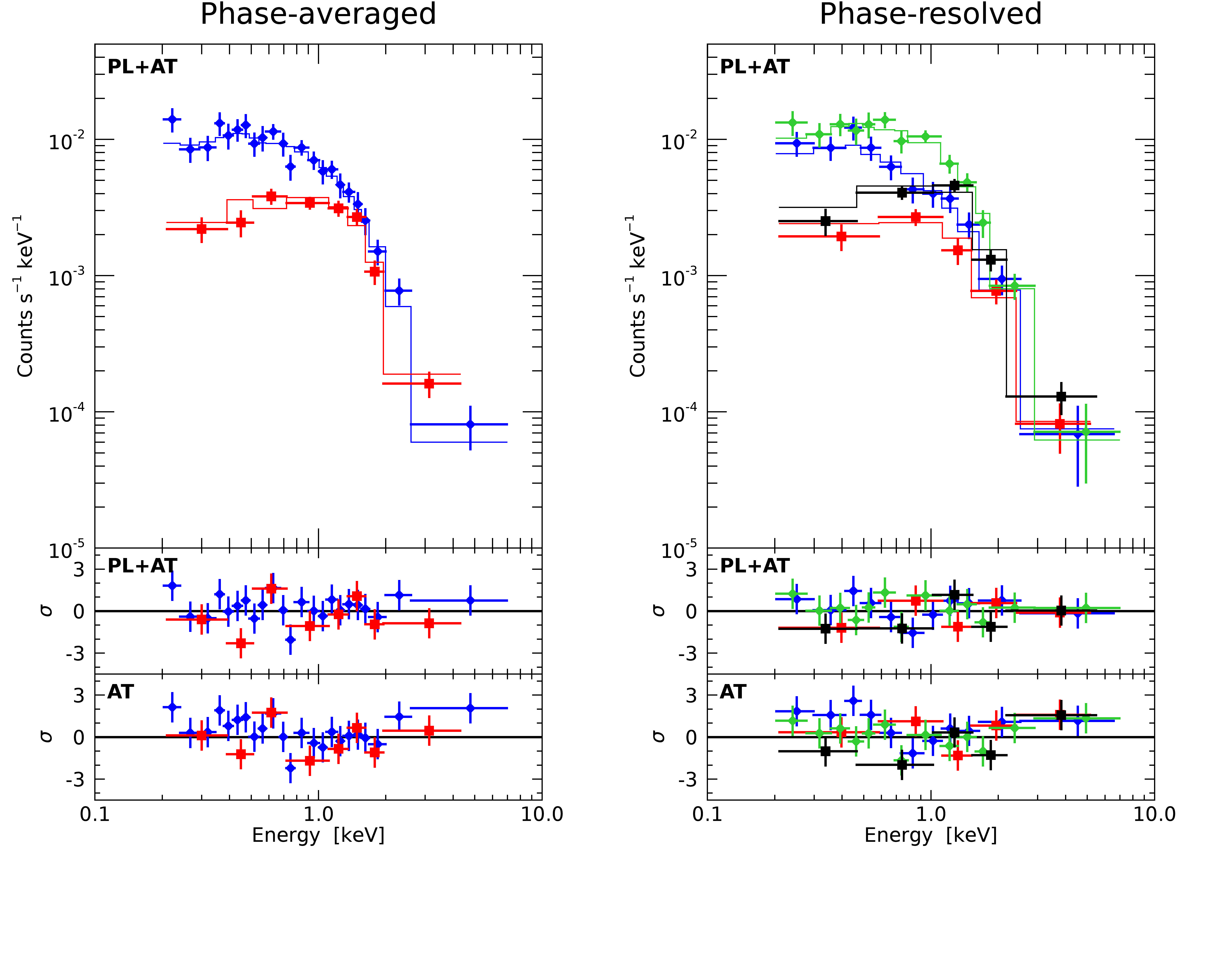}
    \caption{EPIC-pn (diamonds) and -MOS (squares) X-ray spectra of \psrx in the Q-mode. Left panel: phase-averaged spectra (blue: pn, red: MOS); right panel: phase-resolved spectra,  where the spectra at the minimum phase are in blue (pn) and red (MOS), while those at  maximum phase are in green (pn) and black (MOS). The top panels show the best fit of the case $\xi=5^\circ$ and $\chi=3^\circ$ using an absorbed power law plus hydrogen atmosphere model (PL+AT); the corresponding residuals in units of $\sigma$ are shown in the middle panels. The lower panels show  the residuals of the best-fitting atmosphere model (AT). While the fits of the phase-averaged spectra with the two models are equally good, the addition of the power law component significantly improves the fit of the phase-resolved spectra.}
    \label{fig:specAT}
\end{figure*}

For what concerns the B-mode, the atmosphere model alone can fit well the phase-resolved spectra, but only in the configurations with large $\xi$ and $\chi$ (nhp $>0.47$).  
With the addition of a power law (with fixed $\Gamma=2.3$, due to the lower statistics in the B-mode) 
all the geometrical configurations give acceptable spectral fits (nhp $>0.30$) and light curves (KS-test probability $> 0.20$). The latter are shown in Figure \ref{fig:lc_50_B}. The inferred magnetic field, although with large uncertainties, is  lower than that found for the Q-mode. For  the best fitting geometry of the Q-mode ($\xi=5^\circ$, $\chi=3^\circ$), we get $kT=0.082_{-0.009}^{+0.003}$ keV, $R_{\rm cap}=170_{-25}^{+35}$ m and $B=(2.0_{-0.6}^{+1.0}) \times 10^{12}$ G (see Table~\ref{tab:model} for all the details).

\begin{figure*}[htbp!]
    \centering
    \includegraphics[height=9cm]{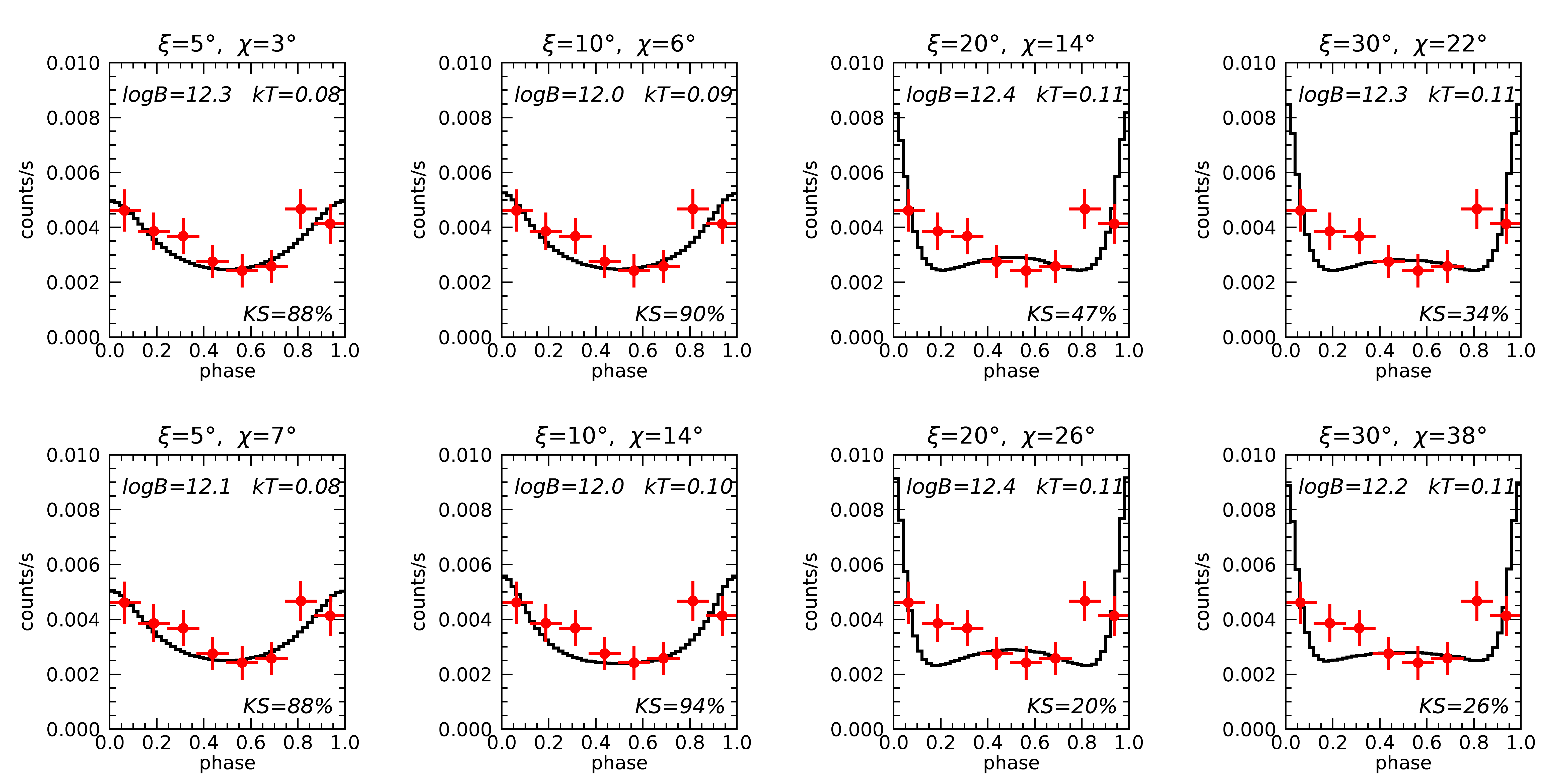}
    \caption{Same as Figure \ref{fig:lc_50}, but for the B-mode.}
    \label{fig:lc_50_B}
\end{figure*}

\subsection{Condensed magnetized surface}
 \label{sec:solid}
 
In the presence of a strong magnetic field, the NS surface can undergo a phase transition into a condensed state, if its temperature is below a critical value which is a function of the magnetic field strength and composition \citep[ and references therein]{med07}.
A high magnetic field strongly affects the properties of atoms, molecules and plasma. This may lead to the formation of linear molecular chains aligned with the magnetic field which can then form a condensate via covalent bonding. The critical temperature at which condensation occurs, at a given $B$, depends on the chemical composition. Higher $Z$ elements form a condensate at lower temperatures with respect to pure hydrogen.

If the polar caps have a temperature of hundreds of eV, a field of $\sim\!10^{14}$ G is needed for magnetic condensation to occur. This value is far above the dipolar magnetic field of \psrx, but we can not exclude the possibility that this pulsar has such a high magnetic field if multipolar components are present close to the polar caps. 

\citet{pot12} developed a simple analytical expression for the emissivity from a condensed iron surface with two extreme approximations for the response of ions to electromagnetic waves: one neglects the Coulomb interactions between ions (\textit{free} case), while the other treats ions as frozen at their equilibrium positions in the Coulomb lattice (\textit{fixed} case). The true radiation properties of a condensed magnetized surface should be in-between these limits (see the discussion in \citealt{tur04}).

The condensed surface emits radiation with monochromatic intensity $I_\nu = \varepsilon_\nu B_\nu$, where $B_\nu$ is the Planck  spectral radiance and $\varepsilon_\nu$ is the dimensionless emissivity, which depends on the magnetic field intensity and also on $\theta_B$, $\theta_k$ and $\phi_k$. As mentioned in section \ref{sec:model}, for nearly point-like magnetic polar caps $\theta_B=0^\circ$ and the emissivity has  azimuthal symmetry.

\begin{figure}[htbp!]
\centering
  \includegraphics[width=0.9\linewidth]{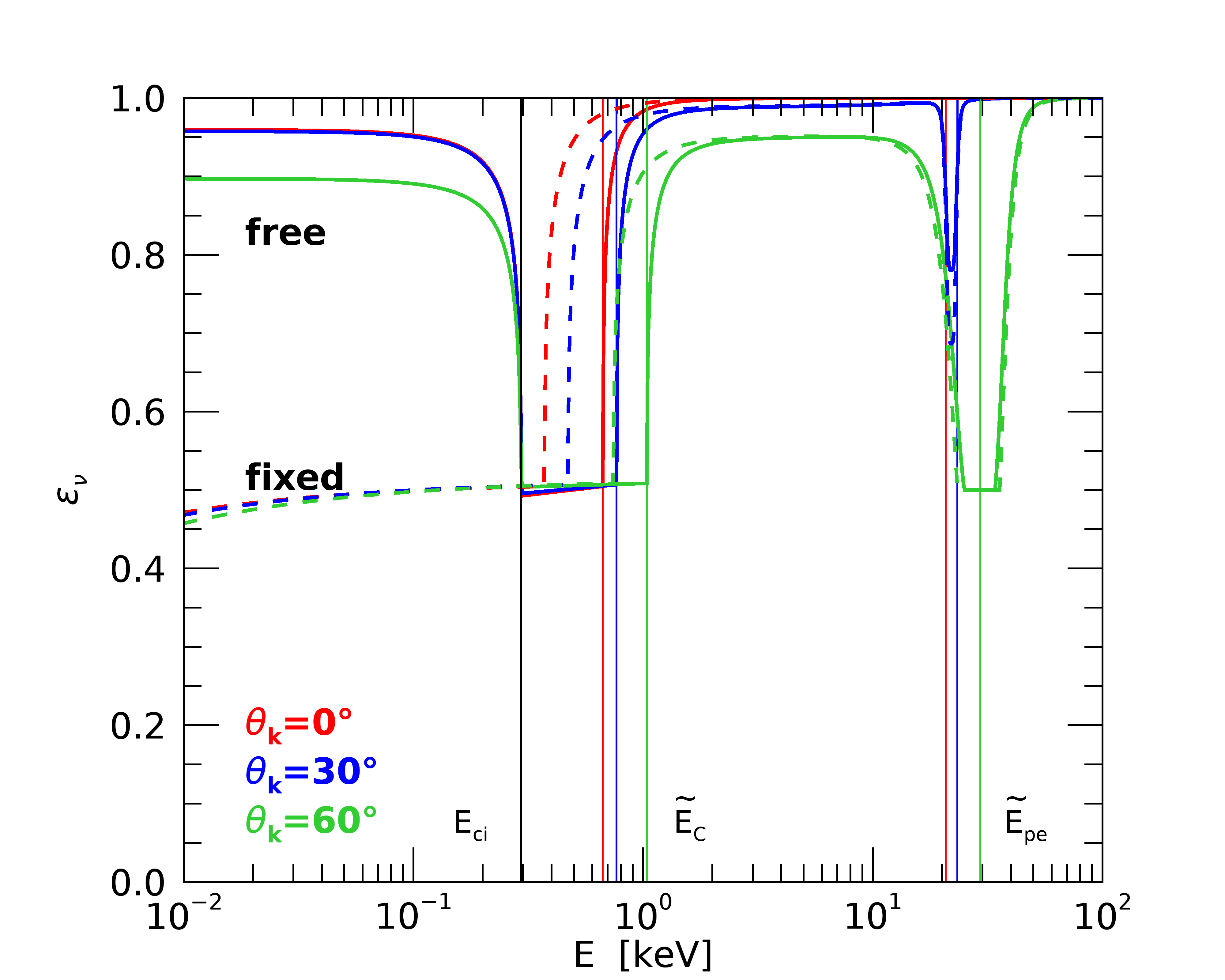}
  \caption{Emissivity as a function of photon energy for a condensed iron surface (free ions, solid lines; fixed ions, dashed lines) for a magnetic field $B=10^{14}$ G normal to the surface ($\theta_B=0^\circ$) and $\rho = 1.1 \times 10^6$ g cm$^{-3}$, see equation \ref{eq:rho}. 
The red lines correspond to an incident angle $\theta_k=0^\circ$, the blue ones to $\theta_k=30^\circ$ and the green ones to $\theta_k=60^\circ$. The vertical lines indicate the values of  $E_{\rm ci}$, $\tilde{E}_{\rm C}$ and $\tilde{E}_{\rm pe}$ (see equations \ref{eq:ece} -- \ref{eq:ec}).}
  \label{fig:emissivity}
\end{figure}

The emissivities  in the free and the fixed cases, computed with the analytic formulae of \citet{pot12}, are plotted in Figure~\ref{fig:emissivity} for $B=10^{14}$~G, $\theta_B=0^\circ$, and three different incident angles ($\theta_k=0^\circ$, $30^\circ$ and $60^\circ$).
The characteristic energies that appear in the plot are: 
the ion cyclotron energy $E_{\rm ci}$,  $\tilde{E}_{\rm C}$, and $\tilde{E}_{\rm pe}$. The latter are functions of the ion-electron cyclotron energies, the electron plasma energy and $\theta_k$:
\begin{equation}
E_{\rm ce} = \hbar e B/m_e c = 115.77~B_{13}~\mathrm{keV}
\label{eq:ece}
\end{equation}

\begin{equation}
E_{\rm ci} = \hbar Z e B/A m_u c = 0.0635~(Z/A)~B_{13}~\mathrm{keV}
\label{eq:eci}
\end{equation}

\begin{equation}
E_{\rm pe} = (4 \pi \hbar^2 e^2 n_e/m_e)^{1/2} = 0.028 \sqrt{\rho Z/A}~\mathrm{keV} 
\label{eq:epe}
\end{equation}

\begin{equation}
\tilde{E}_{\rm pe} = E_{\rm pe} \sqrt{3 - 2 \cos{\theta_k}}
\label{eq:epet}
\end{equation}

\begin{equation}
\tilde{E}_{\rm C} = E_{\rm ci} + \tilde{E}_{\rm pe}^2/E_{\rm ce}.
\label{eq:ec}
\end{equation}

\noindent $\rho$ of equation \ref{eq:epe} is the density at the surface in g cm$^{-3}$,

\begin{equation}
\rho \approx 8.9 \times 10^{3} A Z^{-0.6} B_{13}^{1.2}~\mathrm{g~cm}^{-3},
\label{eq:rho}
\end{equation}
\noindent see \citet{lai01}.

In the free ions case, the emissivity $\varepsilon_\nu$ exhibits different behaviors in three characteristic energy ranges, that depend on the magnetic field intensity and on the incident angle. The fixed case is significantly different from the free one only at low energies, $\lesssim \tilde{E}_{\rm C}$. The weak dependence of $\varepsilon_\nu$ on the temperature has been neglected (as the temperature increases, the transitions of $\varepsilon_\nu$ between characteristic energy ranges become smoother). The bulk of the calculations employed in the fitting was done at $T_\mathrm{eff} = 1$~MK, but a change up to a factor 3 affects the results by an amount similar to the typical error in the fits, as it is discussed in \citet{pot12}.
Figure \ref{fig:emissivity} shows that magnetic beaming of the emission from a condensed surface arises only for photon energies close to the characteristic energies given above, and it grows as $\theta_k$ increases.

We computed the spectra and pulse profiles produced by a hot polar cap with a condensed iron surface, using our baseline values for the geometry of \psrx (Table \ref{tab:angles}), and different values of $kT$ and $B$ in the appropriate range to have a condensed surface. As previously done with the hydrogen atmosphere, for each set of parameters we produced a model of the phase-averaged spectrum and implemented it in XSPEC. 
We verified that for all the pairs of $\xi$ and $\chi$ in Table~\ref{tab:angles} it was possible to find an acceptable spectral fit, but only the most-misaligned geometries  give rise to a light curve pulsed enough. Therefore, in the following we consider only the most favorable case,  $\xi=30^\circ$ and $\chi=38^\circ$.

The phase-averaged spectrum of the Q-mode could be fitted only with the addition of a non-thermal component, but it is impossible to reproduce the observed pulsed fraction because in any case the power law component adds more unpulsed counts. 
Similar results were found in the analysis of the B-mode data.

\setlength{\tabcolsep}{1em}
\begin{table*}[htbp!]
\centering \caption{Results for the phase-resolved spectra of Q- and B-modes}
\label{tab:model}
\begin{tabular}{lcccccc}
\toprule
				& Q-mode			& Q-mode			& Q-mode		& B-mode			& B-mode			& B-mode		\\[5pt]

				& H atmosphere			& Free Ions			& Fixed Ions		& H atmosphere			& Free Ions			& Fixed Ions		\\[5pt]
\midrule\\[-5pt]
$\xi~{}^{\rm a}$ ($^\circ$)	& $5$				& $30$				& $30$			& $5$				& $30$				& $30$			\\[5pt]
$\chi~{}^{\rm a}$ ($^\circ$)	& $3$				& $38$				& $38$			& $3$				& $38$				& $38$			\\[5pt]
\midrule\\[-5pt]
$\Gamma$			& $2.5 \pm 0.2$  		& $2.4 \pm 0.2$  		& $2.5_{-0.1}^{+0.2}$	& $2.3~^{\rm a}$  		& $2.3~^{\rm a}$  		& $2.3~^{\rm a}$	\\[5pt]
$K_{\rm min}~^{\rm b}$		& $2.3_{-0.4}^{+0.3}$  		& $1.8 \pm 0.5$  		& $2.2_{-0.2}^{+0.4}$	& $1.0 \pm 0.3$   		& $0.8 \pm 0.3$			& $1.0 \pm 0.2$		\\[5pt]
$K_{\rm max}~^{\rm b}$		& $2.3~^{\rm a}$  		& $3.7_{-0.6}^{+0.3}$ 		& $4.0_{-0.2}^{+0.5}$ 	& $1.0~^{\rm a}$		& $1.5 \pm 0.3$   		& $1.5 \pm 0.3$		\\[5pt]
$\rm  PF_{PL}~^{\rm c}$		& $0.0$  			& $0.55 \pm 0.11$		& $0.46 \pm 0.08$	& $0.0$   			& $0.43 \pm 0.16$		& $0.30 \pm 0.12$ 	\\[5pt]
\midrule\\[-5pt]
$kT$ (keV)			& $0.089_{-0.005}^{+0.014}$  	& $0.23 \pm 0.02$	  	& $0.24 \pm 0.02$	& $0.082_{-0.009}^{+0.003}$  	& $0.22_{-0.01}^{+0.04}$	& $0.20 \pm 0.02$	\\[5pt]
$R_{\rm cap}~^{\rm d}$ (m) 	& $260_{-60}^{+70}$		& $60 \pm 17$			& $55_{-15}^{+5}$	& $170_{-25}^{+45}$		& $38_{-9}^{+7}$		& $60_{-10}^{+20}$	\\[5pt]
$B$ (G)				& $(4.0_{-0.7}^{+0.9})\x10^{12}$& $(1.8 \pm 0.3)\x10^{14}$	& $>6\x10^{15}$  	& $(2.0_{-0.6}^{+1.0})\x10^{12}$& $[1,6]\x10^{14}$	 	& $>8\x10^{15}$		\\[5pt]
\midrule\\[-5pt]

$\rm Flux_{min}^{0.5-2}$	& $8.9 \pm 0.5$  		& $9.3 \pm 0.5$			& $9.2 \pm 0.5$		& $4.4_{-0.3}^{+0.4}$  		& $4.6 \pm 0.3$			& $4.5 \pm 0.3$		\\[5pt]
$\rm Flux_{max}^{0.5-2}$	& $16.7 \pm 0.6$  		& $17 \pm 1$			& $19 \pm 1$		& $6.9 \pm 0.4$   		& $6.9_{-0.6}^{+0.4}$		& $6.8 \pm 0.4$		\\[5pt]

$\chi_{\nu}^2$/dof		& 0.82/29			& 1.20/28			& 1.06/28		& 1.15/26			& 1.08/25			& 1.13/25		\\[5pt]
nhp				& 0.73				& 0.22				& 0.38			& 0.27				& 0.36				& 0.30			\\[5pt]
KS$~^{\rm e}$   		& 0.31				& 0.46				& 0.49			& 0.88				& 0.93				& 0.87			\\[5pt]

\bottomrule
\end{tabular}

\raggedright
Joint fits of pn + MOS phase-resolved spectra with magnetized hydrogen atmosphere and condensed magnetized surface models.
\nh is fixed to $4.3 \times 10^{20}$ cm$^{-2}$.
The fluxes, corrected for the absorption, are in units of $10^{-15}$~erg~cm$^{-2}$~s$^{-1}$.
Errors and upper limits at $1\sigma$.\\
$^{\rm a}$ Fixed value.\\
$^{\rm b}$ Normalization of the power law at 1 keV in units of $10^{-6}$~photons~cm$^{-2}$~s$^{-1}$~keV$^{-1}$.\\
$^{\rm c}$ Pulsed fraction of the power law component assuming a sinusoidal modulation.\\
$^{\rm d}$ Radius of the cap for an assumed distance of 0.89 kpc.\\
$^{\rm e}$ KS probability for the observed pulse profile. \\
\end{table*}

In conclusion, the condensed surface emission model requires the presence of an additional power law component in the spectrum and, 
in order to reproduce the observed pulsed fraction, also this non-thermal component has to be pulsed.
Not surprisingly, this is similar to the case of blackbody thermal emission examined in section~\ref{sec:bb}.
To quantify the required modulation of the non-thermal component, we performed phase-resolved spectroscopy using the two phase intervals defined above.  We fitted with the condensed surface model plus a power law with normalization free to vary between the pulse maximum ($K_{\rm max}$) and  minimum ($K_{\rm min}$). 
The best fit parameters are summarized in Table \ref{tab:model}, where also the pulsed fraction of the power law is reported.

\bigskip

\section{Discussion}
\label{sec:disc}

Most previous analyses of the X-ray emission from the radio and X-ray mode-switching pulsar \psrx \citep{zha05,her13,mer13,mer16}, with the notable exception of that in \citet{sto14}, were based on simple combinations of blackbody and power law components to model the mix of thermal and non-thermal emission detected in such an old pulsar.
The purpose of our work was to explore in a more quantitative way realistic models for the thermal emission of \psrx, exploiting all the available X-ray data and taking into account the most recent (and less constraining) geometrical configurations derived from the radio observations of this pulsar. In fact, compared to brighter X-ray pulsars that can provide spectra with a better statistics, \psrx has the advantage of a rather well known geometry. This is of great importance since it reduces the number of parameters (or at least their allowed ranges) on which the spectral and timing properties depend.
In this work we adopted a  value of the neutron star mass to radius ratio $M/R$ = $1.5\msun / 12~\mathrm{km} = 0.125 \msun/\mathrm{km}$. With lower values of the compactness it is possible to obtain higher pulsed fractions (for example a compactness of 0.1 $\msun/\mathrm{km}$ was used by \citealt{sto14}).

We could adequately fit the 2011 plus 2014 X-ray spectra of  \psrx during both radio modes with the sum of a power law and a blackbody, confirming the results already reported with the individual data sets of the two observing campaigns \citep{her13,mer16}. In addition, the higher statistics provided by the combined data disfavors the fit of the B-mode spectrum with a single blackbody, supporting the presence of thermal and non-thermal components in both modes, as proposed by these authors on the basis of their spectral-timing analysis of the pulsed and unpulsed emission.
 
It is natural to associate the pulsed thermal component to the polar caps, but since the geometric configuration of \psrx implies that the emitting polar region is visible at all rotational phases, it is impossible to reproduce the large and energy-dependent (in the Q-mode) pulsed fraction unless the thermal emission is magnetically beamed \citep{sto14}.  A further problem of fitting with simple blackbody models is that, for the geometry of \psrx, even the emission from a hot polar cap itself supplies a significant amount of unpulsed flux, about 5 times brighter than the pulsed one. This is at variance with the results of previous analyses (confirmed here, see Table~\ref{tab:ML3D}) showing that a single power law is adequate to fit the unpulsed emission \citep{her13,mer16}.

These problems can not be solved adopting a model of thermal emission from polar caps with a condensed iron surface, as it could be expected in the presence of strong multipolar magnetic field componentsthat could give  a field higher up to orders of magnitude than the dipole. However, values as large as  $\sim\!10^{16}$ G, as obtained in the fixed ions case, seem rather unrealistic.
Although the condensed iron surface model can fit well the phase-averaged spectra of both the Q- and B-mode, we faced the same problems found with the blackbody, even in the most misaligned configuration consistent with the radio data ($\xi=30^\circ$, $\chi=38^\circ$).
The observed pulse profiles can be reproduced only if we add in the fits a non-thermal power law emission significantly pulsed (see Table~\ref{tab:model}). 

For the case of a magnetized, partially ionized hydrogen atmosphere, acceptable fits to the phase-averaged spectra of the Q- and B-mode could be found for all the geometrical configurations derived from the radio data. 
If only the phase-averaged spectra are considered, the fits with a hydrogen atmosphere are acceptable without the need of an additional power law component, contrary to the case of blackbody and condensed surface models. 
In this case, the absence of the power law contribution at low energies leads to small values of interstellar absorption.  
The best fit temperatures, in the range $kT = 0.10 -0.15$ keV, are lower than those obtained with the blackbody model, as it is always the case when hydrogen model atmospheres are applied (e.g., \citealt{zav04}).
Correspondingly, the emitting radii are larger and  compatible with the expected size of the magnetic polar cap for a dipolar field, $R_{\rm PC}=(2 \pi R^3/P c)^{1/2} \approx 180$ m (for $R = 12$~km). 
Compared to the results of \citet{sto14} for the Q-mode, we find slightly lower best fit temperatures and larger emitting radii (even accounting for the different distance used by these authors). Details of the models, as well as different assumptions for the star compactness, viewing geometry and magnetic field, could possibly explain this discrepancy. 

However, to fit the phase-resolved spectra an unpulsed power law component is required in addition to the emission from the polar caps modeled with the magnetized hydrogen atmosphere. We found acceptable fits for all the considered geometrical configurations, even if the more misaligned ones are disfavoured. In fact, for large $\xi$, we expect to see also the fan-beamed emission, that gives rise to a second peak in the light curve, at phase $0.5$. The higher the magnetic field, the more intense the peak is.  However, in the pulse profile of \psrx this second peak is absent and this explains why the magnetic field derived with the fit is the lowest allowed in our grid. We do not expect that the match between the observed and the simulated pulse profile would improve for lower values of $B$ that, moreover, would be inconsistent with the dipole field derived from the timing parameters of \psrx.

On the other hand, more aligned configurations predict a light curve with only one broad peak at phase $0$, similar to the observed one. In fact, the best-fitting configuration for the Q-mode has $\xi=5^\circ$ and $\chi=3^\circ$, and the spectral parameters of the thermal component are $kT\approx0.09$ keV, $R_{\rm cap} \approx 260$ m and $B\approx 4 \times 10^{12}$ G. Remarkably, the magnetic field is fully compatible with the value at the poles derived in the dipole approximation.

\begin{figure}[htbp!]
    \centering
     \includegraphics[width=0.9\linewidth]{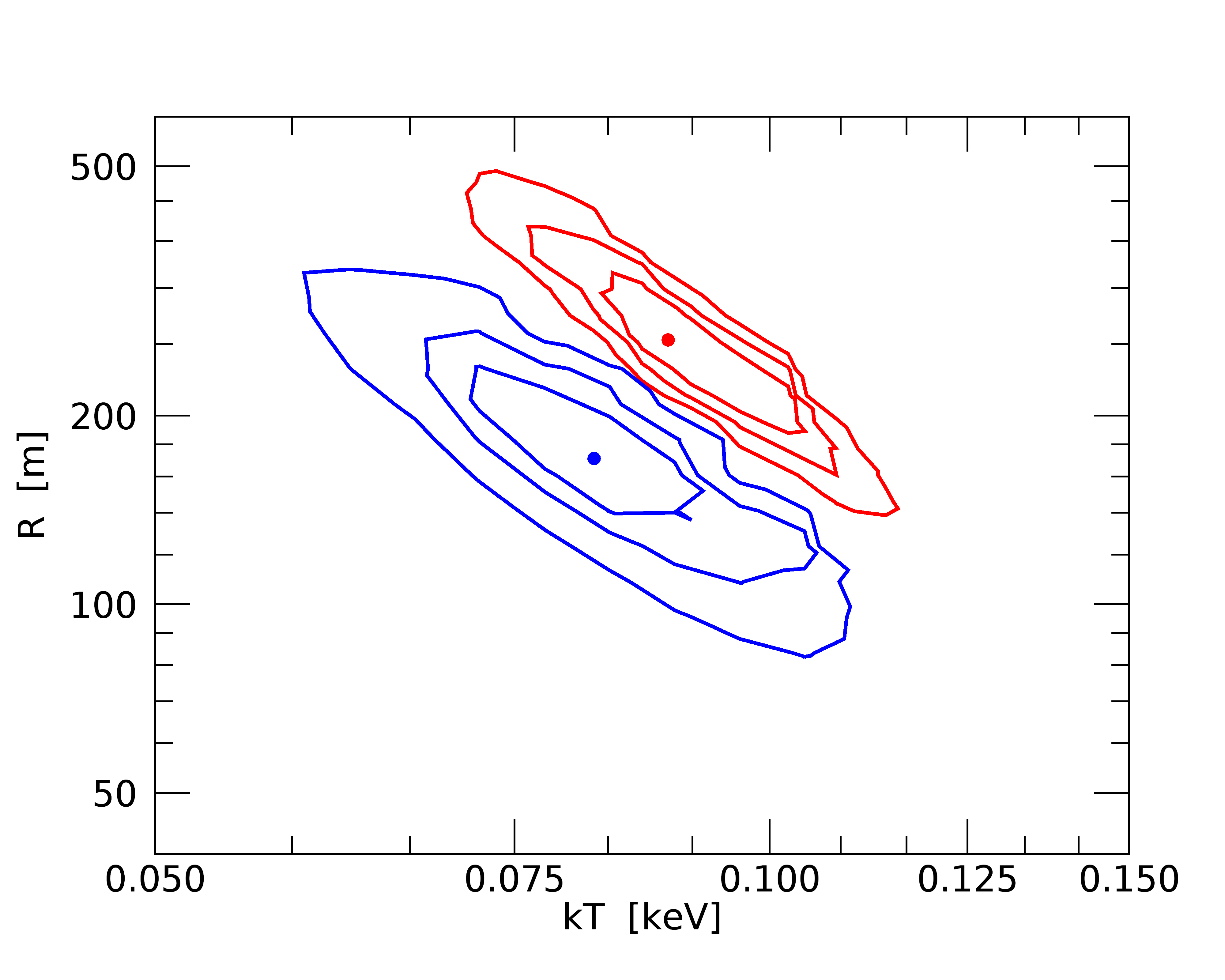}
    \caption{Confidence regions (1, 2 and 3 $\sigma$ confidence levels) of the polar cap temperature and radius when \psrx is in the Q- (red lines) and in the B-mode (blue lines). The spectral parameters are derived using a magnetized hydrogen atmosphere model.}
    \label{fig:con}
\end{figure}

The B-mode has a similar, but less pulsed, light curve. Owing to the lower counting statistics (see Table~\ref{tab:obs}), all the explored geometrical configurations give acceptable spectra and light curves, but, using the same argument we put forward for the Q-mode, we tend to exclude the more misaligned configurations. For the most favoured geometrical configuration of the Q-mode, the best-fit B-mode parameters are $kT\approx0.08$ keV, $R_{\rm cap} \approx 170$ m and $B\approx 2 \times 10^{12}$ G. The confidence regions of $R_{\rm cap}$ and $kT$ for the two radio modes are shown in Figure \ref{fig:con}.  It is clear that, with the current data it is impossible to ascertain whether the flux difference between the two modes is due to a change in the temperature or in the size of the  emitting area.

\begin{figure}[htbp!]
    \centering
     \includegraphics[width=0.9\linewidth]{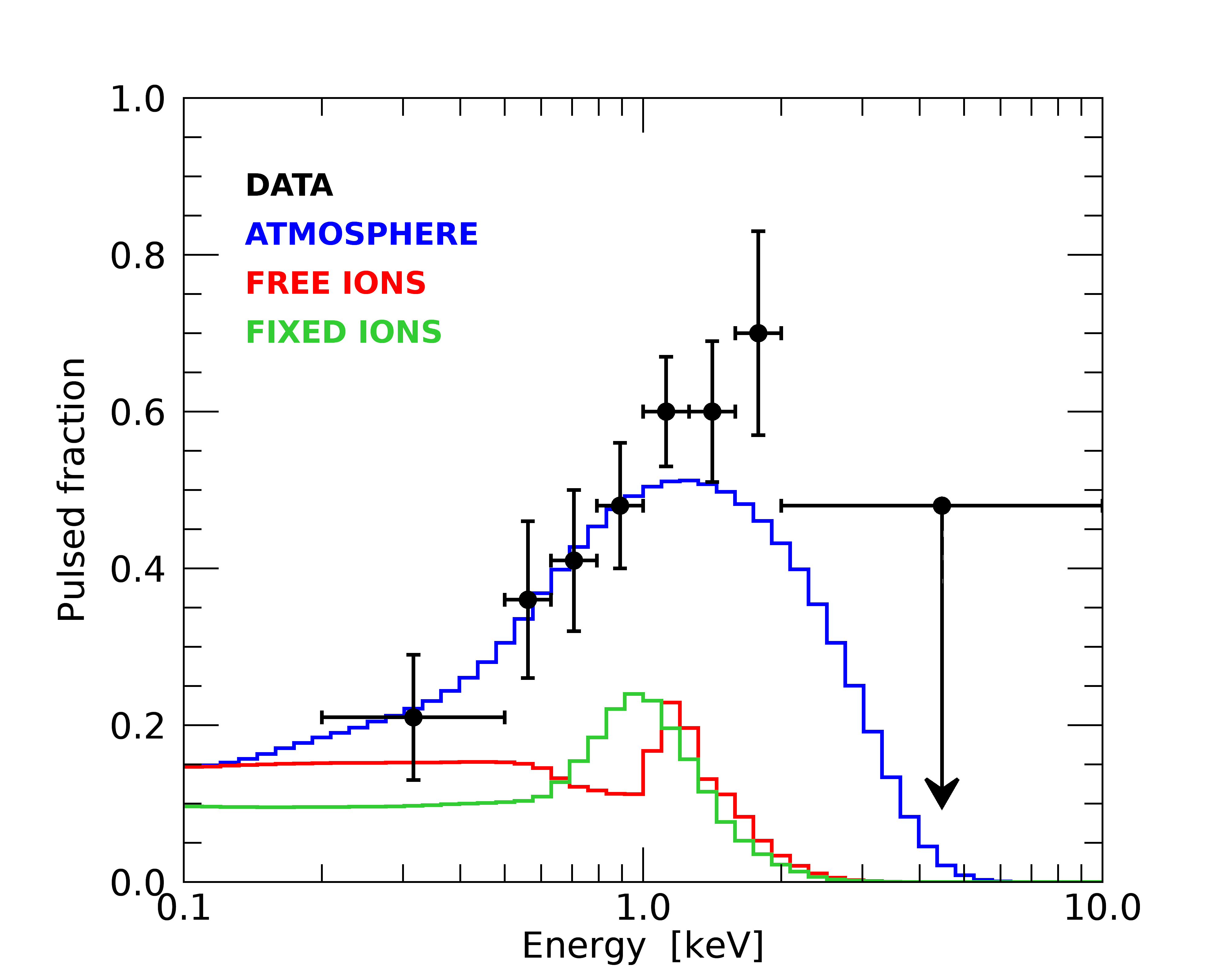}
    \caption{Pulsed fraction as a function of energy computed for the case of a hydrogen atmosphere with $\xi=5^\circ$ and $\chi=3^\circ$ (blue line) and of a condensed iron surface with $\xi=30^\circ$ and $\chi=38^\circ$ in the free (red) and fixed (green) ions approximations. Differently to the atmosphere model, in the condensed surface case the plotted lines refer only to the thermal contribution from the polar caps. The addition of an unpulsed power law component would reduce the pulsed fraction. The black dots with error bars indicate the observed pulsed fraction of the Q-mode.}
    \label{fig:pf_models}
\end{figure}

The hydrogen atmosphere model can also provide a better match with the observed pulsed fraction of \psrx than the blackbody and condensed surface models, see Figure \ref{fig:pf_models}. A similar result was recently reported for PSR$\,$B0823$+$26, another mode switching pulsar with a mainly thermal spectrum and a large pulsed fraction \citep{her18}. Remarkably, the best fitting hydrogen atmosphere model for PSR$\,$B0823$+$26 was obtained for a geometrical configuration ($\xi=81^\circ$ and $\chi=66^\circ$) different from that derived from the radio data ($\xi=81^\circ$ and $\chi=84^\circ$).

\section{Conclusions}

Our spectral analysis of the summed 2011 and 2014 data of \psrx using blackbody and power law models confirmed the results obtained in their separate analysis \citep{her13,mer16}. In addition, thanks to the increased counting statistics, we showed that the single blackbody fit to the B-mode emission, that was consistent with the 2014 data, is actually disfavored. However, if we adopt the pulsar geometrical configuration derived from recent radio studies \citep{bil18}, these simple models face two problems: \textit{a)} they can not give rise to the significant and energy-dependent pulsed fraction observed in the Q-mode; \textit{b)} they can not correctly reproduce the spectrum of the unpulsed flux, that should  be dominated by the blackbody emission, while it can be fitted well by a single power law.

Replacing the blackbody with a model of thermal emission from an iron condensed surface, as it could be expected in the presence of strong multipolar components of the magnetic field, can not solve these problems. Therefore, one has to invoke that, in the blackbody and solid surface case, also the non-thermal emission is, to some extent, pulsed. Such a possibility is plausible, as shown by theoretical expectations and observations of other pulsars, and it is consistent with the current data of \psrx.   

A  good description of the \psrx spectra and pulse profiles could be obtained using a magnetized hydrogen atmosphere model plus an unpulsed non-thermal component. In this case it was possible to fit well the  phase-resolved spectra of both modes, for several geometrical configurations consistent with the radio data. As first shown by \citet{sto14} for this pulsar, the significant beaming of the emission predicted by magnetized atmosphere models gives rise to pulse profiles more consistent with the observed ones, and, as typically observed with atmosphere models, yields smaller temperatures and larger emission radii than those  of blackbody fits ($kT\sim0.09$ keV, $R\sim260$ m, for the Q-mode; $kT\sim0.08$ keV,  $R\sim170$ m, for the B-mode). 
We explored a few representative geometrical configurations derived from the radio data and found that, for surface magnetic fields of the order of the dipole value derived from the timing parameters, the most aligned configurations (i.e. $\xi=5^\circ$ and $\chi=3^\circ$) are favoured.

The results reported here underline the importance of exploiting the full spectral and timing information in the analysis of X-ray pulsars and the strong constraints posed on the models by the knowledge of the pulsar geometry,  which unfortunately is not available for most pulsars.

\begin{acknowledgements}
We are grateful to an anonymous referee for constructive suggestions and to George Pavlov for several useful comments. 
We acknowledge financial contribution from the agreement ASI-INAF n.2017-14-H.0. Part of this work has been funded using resources from the research grant “iPeska” (P.I. Andrea Possenti) funded under the INAF national call Prin-SKA/CTA approved with the Presidential Decree 70/2016.
This work is based on observations obtained with \xmm, an European Space Agency (ESA) science mission with instruments and contributions directly funded by ESA Member States and NASA. 
VS thanks Deutsche Forschungsgemeinschaft (DFG) for financial support (grant WE 1312/51-1). His work was also supported by the grant 14.W03.31.0021 of the Ministry of Education and Science of the Russian Federation.
\end{acknowledgements}

\bibliographystyle{aa}
\bibliography{bibliography}

\begin{thebibliography}{56}
\expandafter\ifx\csname natexlab\endcsname\relax\def\natexlab#1{#1}\fi

\bibitem[{{Arons} \& {Scharlemann}(1979)}]{aro79}
{Arons}, J. \& {Scharlemann}, E.~T. 1979, \apj, 231, 854

\bibitem[{{Arras} {et~al.}(2004){Arras}, {Cumming}, \& {Thompson}}]{arr04}
{Arras}, P., {Cumming}, A., \& {Thompson}, C. 2004, \apjl, 608, L49

\bibitem[{{Beloborodov}(2002)}]{bel02}
{Beloborodov}, A.~M. 2002, \apjl, 566, L85

\bibitem[{{Bilous}(2018)}]{bil18}
{Bilous}, A.~V. 2018, \aap, 616, A119

\bibitem[{{Bilous} {et~al.}(2016){Bilous}, {Kondratiev}, {Kramer}, {Keane},
  {Hessels}, {Stappers}, {Malofeev}, {Sobey}, {Breton}, {Cooper}, {Falcke},
  {Karastergiou}, {Michilli}, {Os{\l}owski}, {Sanidas}, {ter Veen}, {van
  Leeuwen}, {Verbiest}, {Weltevrede}, {Zarka}, {Grie{\ss}meier}, {Serylak},
  {Bell}, {Broderick}, {Eisl{\"o}ffel}, {Markoff}, \& {Rowlinson}}]{bil16}
{Bilous}, A.~V., {Kondratiev}, V.~I., {Kramer}, M., {et~al.} 2016, \aap, 591,
  A134

\bibitem[{{Cooper} \& {Kaplan}(2010)}]{coo10}
{Cooper}, R.~L. \& {Kaplan}, D.~L. 2010, \apjl, 708, L80

\bibitem[{{Cordes}(2013)}]{cor13}
{Cordes}, J.~M. 2013, \apj, 775, 47

\bibitem[{{Cordes} \& {Lazio}(2002)}]{cor02}
{Cordes}, J.~M. \& {Lazio}, T.~J.~W. 2002, ArXiv: 0207156

\bibitem[{{Deshpande} \& {Rankin}(2001)}]{des01}
{Deshpande}, A.~A. \& {Rankin}, J.~M. 2001, \mnras, 322, 438

\bibitem[{{Gil} {et~al.}(1984){Gil}, {Gronkowski}, \& {Rudnicki}}]{gil84}
{Gil}, J., {Gronkowski}, P., \& {Rudnicki}, W. 1984, \aap, 132, 312

\bibitem[{{Ginzburg}(1970)}]{gin70}
{Ginzburg}, V.~L. 1970, {The propagation of electromagnetic waves in plasmas}
  ({Pergamon Press, Oxford})

\bibitem[{{Greenstein} \& {Hartke}(1983)}]{gre83}
{Greenstein}, G. \& {Hartke}, G.~J. 1983, \apj, 271, 283

\bibitem[{{Harding} \& {Muslimov}(2001)}]{har01}
{Harding}, A.~K. \& {Muslimov}, A.~G. 2001, \apj, 556, 987

\bibitem[{{Harding} \& {Muslimov}(2002)}]{har02a}
---. 2002, \apj, 568, 862

\bibitem[{{Hermsen} {et~al.}(2018){Hermsen}, {Kuiper}, {Basu}, {Hessels},
  {Mitra}, {Rankin}, {Stappers}, {Wright}, {Grie{\ss}meier}, {Serylak},
  {Horneffer}, {Tiburzi}, \& {Ho}}]{her18}
{Hermsen}, W., {Kuiper}, L., {Basu}, R., {et~al.} 2018, \mnras, 480, 3655

\bibitem[{{Hermsen} {et~al.}(2017){Hermsen}, {Kuiper}, {Hessels}, {Mitra},
  {Rankin}, {Stappers}, {Wright}, {Basu}, {Szary}, \& {van Leeuwen}}]{her17}
{Hermsen}, W., {Kuiper}, L., {Hessels}, J.~W.~T., {et~al.} 2017, \mnras, 466,
  1688

\bibitem[{{Hermsen} {et~al.}(2013)}]{her13}
{Hermsen}, W. {et~al.} 2013, Science, 339, 436

\bibitem[{{Kaminker} {et~al.}(2014){Kaminker}, {Kaurov}, {Potekhin}, \&
  {Yakovlev}}]{kam14}
{Kaminker}, A.~D., {Kaurov}, A.~A., {Potekhin}, A.~Y., \& {Yakovlev}, D.~G.
  2014, \mnras, 442, 3484

\bibitem[{{Lai}(2001)}]{lai01}
{Lai}, D. 2001, Rev. Mod. Phys., 73, 629

\bibitem[{{Malofeev} {et~al.}(2000){Malofeev}, {Malov}, \&
  {Shchegoleva}}]{mal00}
{Malofeev}, V.~M., {Malov}, O.~I., \& {Shchegoleva}, N.~V. 2000, Astronomy
  Reports, 44, 436

\bibitem[{{Medin} \& {Lai}(2007)}]{med07}
{Medin}, Z. \& {Lai}, D. 2007, \mnras, 382, 1833

\bibitem[{{Mereghetti} {et~al.}(2016){Mereghetti}, {Kuiper}, {Tiengo},
  {Hessels}, {Hermsen}, {Stovall}, {Possenti}, {Rankin}, {Esposito}, {Turolla},
  {Mitra}, {Wright}, {Stappers}, {Horneffer}, {Oslowski}, {Serylak}, \&
  {Grie{\ss}meier}}]{mer16}
{Mereghetti}, S., {Kuiper}, L., {Tiengo}, A., {et~al.} 2016, \apj, 831, 21

\bibitem[{{Mereghetti} \& {Rigoselli}(2017)}]{mer17}
{Mereghetti}, S. \& {Rigoselli}, M. 2017, Journal of Astrophysics and
  Astronomy, 38, 54

\bibitem[{{Mereghetti} {et~al.}(2013){Mereghetti}, {Tiengo}, {Esposito}, \&
  {Turolla}}]{mer13}
{Mereghetti}, S., {Tiengo}, A., {Esposito}, P., \& {Turolla}, R. 2013, \mnras,
  435, 2568

\bibitem[{{M{\'e}sz{\'a}ros}(1992)}]{mes92}
{M{\'e}sz{\'a}ros}, P. 1992, {High-energy radiation from magnetized neutron
  stars.} (University of Chicago Press, Chicago)

\bibitem[{{Mitra} \& {Deshpande}(1999)}]{mit99}
{Mitra}, D. \& {Deshpande}, A.~A. 1999, \aap, 346, 906

\bibitem[{{Pavlov} {et~al.}(1994){Pavlov}, {Shibanov}, {Ventura}, \&
  {Zavlin}}]{pav94}
{Pavlov}, G.~G., {Shibanov}, Y.~A., {Ventura}, J., \& {Zavlin}, V.~E. 1994,
  \aap, 289, 837

\bibitem[{{Pechenick} {et~al.}(1983){Pechenick}, {Ftaclas}, \& {Cohen}}]{pec83}
{Pechenick}, K.~R., {Ftaclas}, C., \& {Cohen}, J.~M. 1983, \apj, 274, 846

\bibitem[{{Pons} {et~al.}(2007){Pons}, {Link}, {Miralles}, \&
  {Geppert}}]{pon07}
{Pons}, J.~A., {Link}, B., {Miralles}, J.~A., \& {Geppert}, U. 2007, Physical
  Review Letters, 98, 071101

\bibitem[{{Potekhin}(2014)}]{pot14b}
{Potekhin}, A.~Y. 2014, Physics Uspekhi, 57, 735

\bibitem[{{Potekhin} \& {Chabrier}(2003)}]{pot03}
{Potekhin}, A.~Y. \& {Chabrier}, G. 2003, \apj, 585, 955

\bibitem[{{Potekhin} \& {Chabrier}(2018)}]{pot18}
---. 2018, \aap, 609, A74

\bibitem[{{Potekhin} {et~al.}(2014){Potekhin}, {Chabrier}, \& {Ho}}]{pot14}
{Potekhin}, A.~Y., {Chabrier}, G., \& {Ho}, W.~C.~G. 2014, \aap, 572, A69

\bibitem[{{Potekhin} {et~al.}(2016){Potekhin}, {Ho}, \& {Chabrier}}]{pot16}
{Potekhin}, A.~Y., {Ho}, W.~C.~G., \& {Chabrier}, G. 2016, Proceedings of
  Science, PoS(MPCS2015)016; ArXiv: 1605.01281

\bibitem[{{Potekhin} {et~al.}(2004){Potekhin}, {Lai}, {Chabrier}, \&
  {Ho}}]{pot04}
{Potekhin}, A.~Y., {Lai}, D., {Chabrier}, G., \& {Ho}, W.~C.~G. 2004, \apj,
  612, 1034

\bibitem[{{Potekhin} {et~al.}(2015){Potekhin}, {Pons}, \& {Page}}]{pot15}
{Potekhin}, A.~Y., {Pons}, J.~A., \& {Page}, D. 2015, \ssr, 191, 239

\bibitem[{{Potekhin} {et~al.}(2012){Potekhin}, {Suleimanov}, {van Adelsberg},
  \& {Werner}}]{pot12}
{Potekhin}, A.~Y., {Suleimanov}, V.~F., {van Adelsberg}, M., \& {Werner}, K.
  2012, \aap, 546, A121

\bibitem[{{Rankin}(1993)}]{ran93}
{Rankin}, J.~M. 1993, \apj, 405, 285

\bibitem[{{Rigoselli} \& {Mereghetti}(2018)}]{rig18}
{Rigoselli}, M. \& {Mereghetti}, S. 2018, \aap, 615, A73

\bibitem[{{Romani}(1987)}]{rom87}
{Romani}, R.~W. 1987, \apj, 313, 718

\bibitem[{{Ruderman} \& {Sutherland}(1975)}]{rud75}
{Ruderman}, M.~A. \& {Sutherland}, P.~G. 1975, \apj, 196, 51

\bibitem[{{Shibanov} {et~al.}(1992){Shibanov}, {Zavlin}, {Pavlov}, \&
  {Ventura}}]{shi92}
{Shibanov}, I.~A., {Zavlin}, V.~E., {Pavlov}, G.~G., \& {Ventura}, J. 1992,
  \aap, 266, 313

\bibitem[{{Storch} {et~al.}(2014){Storch}, {Ho}, {Lai}, {Bogdanov}, \&
  {Heinke}}]{sto14}
{Storch}, N.~I., {Ho}, W.~C.~G., {Lai}, D., {Bogdanov}, S., \& {Heinke}, C.~O.
  2014, \apjl, 789, L27

\bibitem[{{Suleimanov} {et~al.}(2009){Suleimanov}, {Potekhin}, \&
  {Werner}}]{sul09}
{Suleimanov}, V., {Potekhin}, A.~Y., \& {Werner}, K. 2009, \aap, 500, 891

\bibitem[{{Suleimanova} \& {Izvekova}(1984)}]{sul84}
{Suleimanova}, S.~A. \& {Izvekova}, V.~A. 1984, \sovast, 28, 32

\bibitem[{{Taverna} {et~al.}(2015){Taverna}, {Turolla}, {Gonzalez Caniulef},
  {Zane}, {Muleri}, \& {Soffitta}}]{tav15}
{Taverna}, R., {Turolla}, R., {Gonzalez Caniulef}, D., {et~al.} 2015, \mnras,
  454, 3254

\bibitem[{{Tsuruta} \& {Cameron}(1966)}]{tsu66}
{Tsuruta}, S. \& {Cameron}, A.~G.~W. 1966, Canadian Journal of Physics, 44,
  1863

\bibitem[{{Turolla} \& {Nobili}(2013)}]{tur13}
{Turolla}, R. \& {Nobili}, L. 2013, \apj, 768, 147

\bibitem[{{Turolla} {et~al.}(2004){Turolla}, {Zane}, \& {Drake}}]{tur04}
{Turolla}, R., {Zane}, S., \& {Drake}, J.~J. 2004, \apj, 603, 265

\bibitem[{{van Adelsberg} \& {Lai}(2006)}]{van06}
{van Adelsberg}, M. \& {Lai}, D. 2006, \mnras, 373, 1495

\bibitem[{{Vigan{\`o}} {et~al.}(2013){Vigan{\`o}}, {Rea}, {Pons}, {Perna},
  {Aguilera}, \& {Miralles}}]{vig13}
{Vigan{\`o}}, D., {Rea}, N., {Pons}, J.~A., {et~al.} 2013, \mnras, 434, 123

\bibitem[{{Yao} {et~al.}(2017){Yao}, {Manchester}, \& {Wang}}]{yao17}
{Yao}, J.~M., {Manchester}, R.~N., \& {Wang}, N. 2017, \apj, 835, 29

\bibitem[{{Zane} \& {Turolla}(2006)}]{zan06}
{Zane}, S. \& {Turolla}, R. 2006, \mnras, 366, 727

\bibitem[{{Zavlin} \& {Pavlov}(2002)}]{zav02}
{Zavlin}, V.~E. \& {Pavlov}, G.~G. 2002, in Neutron Stars, Pulsars, and
  Supernova Remnants, ed. W.~{Becker}, H.~{Lesch}, \& J.~{Tr{\"u}mper}, 263

\bibitem[{{Zavlin} \& {Pavlov}(2004)}]{zav04}
{Zavlin}, V.~E. \& {Pavlov}, G.~G. 2004, \apj, 616, 452

\bibitem[{{Zhang} {et~al.}(2005){Zhang}, {Sanwal}, \& {Pavlov}}]{zha05}
{Zhang}, B., {Sanwal}, D., \& {Pavlov}, G.~G. 2005, \apjl, 624, L109

\end{thebibliography}

\end{document}